\documentclass[usenatbib]{mn2e}
\usepackage{amssymb}
\usepackage{amsmath}
\usepackage{array}
\usepackage{bigstrut}
\usepackage{cellspace}
\usepackage{graphicx}

\newif\ifAMStwofonts
\def\ut #1 #2 { \, \rmn{#1}^{#2}}
\def\grad{\bmath{\nabla}}
\def\cross{\bmath{\times}}
\def\bdot{\bmath{\cdot}}
\def\curl #1 {\grad \cross #1}
\def\div #1 {\grad \bdot #1}
\def\b{\bmath{b}}
\def\v{\bmath{v}}
\def\w{\bmath{w}}
\def\B{\bmath{B}}
\def\E{\bmath{E}}            
\def\e{\bmath{e}}            
\def\we{\bmath{w}_{\rm E}}   
\def\Bh{\bmath{\hat{B}}}
\def\fh{\bmath{\hat{\phi}}}  
\def\rh{\bmath{\hat{r}}}     

\def\rhot{\tilde \rho}     
\def\zt{\tilde z}          
\def\cs{c_{\rm s}}        
\def\beti{\beta_{\rm i 0}}  
\def\bete{\beta_{\rm e 0}}  
\def\vr{v_{r}}               
\def\vf{v_{\phi}}            
\def\vk{v_{\rm K}}           
\def\vz{v_{z}}               

\def\wr{w_{r}}               
\def\wf{w_{\phi}}            
\def\wz{w_{z}}               

\def\Br{B_{r}}
\def\Ba{B_{\phi}}
\def\Bz{B_{z}}

\def\br{b_{r}}
\def\ba{b_{\phi}}
\def\bz{b_{z}}

\def\Er{E_{r}^{\prime}}
\def\Ef{E_{\phi}^{\prime}}
\def\Ez{E_{z}^{\prime}}

\def\er{e_{r}^{\prime}}
\def\ef{e_{\phi}^{\prime}}
\def\ez{e_{z}^{\prime}}

\def\jr{j_{r}}
\def\jf{j_{\phi}}

\def\sigpar{\tilde{\sigma}_{\rm O}}
\def\sigP{\tilde{\sigma}_{\rm P}}
\def\sigH{\tilde{\sigma}_{\rm H}}
\def\etaA{\eta_{\rm A}}
\def\etaO{\eta_{\rm O}}
\def\etaH{\eta_{\rm H}}

\def\Epa{\bmath{E^\prime_\parallel}}  
\def\Epe{\bmath{E^\prime_\perp}}  
\def\J{\bmath{J}}
\def\j{\bmath{j}}

\newcommand{\delt} [1] {\frac{\partial #1}{\partial t}}
\newcommand{\delr} [1] {\frac{\partial #1}{\partial r}}

\newcommand{\delz} [1] {\frac{\partial #1}{\partial z}}
\newcommand{\ee}[1]{\times 10^{#1}}

\usepackage[]{graphics}
\title{Wind-driving protostellar accretion discs. I. Formulation and
parameter constraints} 
\author[A. K\"onigl, R. Salmeron and M. Wardle]
{Arieh K\"onigl$^{1}$\thanks{E-Mail: arieh@jets.uchicago.edu~(AK);\newline
raquel@mso.anu.edu.au~(RS); wardle@physics.mq.edu.au~(MW)}, Raquel
Salmeron$^2$\footnotemark[1] and Mark Wardle$^3$\footnotemark[1] \\
$^1$Department of Astronomy \& Astrophysics and The Enrico Fermi
Institute, The University of Chicago, Chicago IL 60637, USA \\
$^2$Research School of Astronomy \& Astrophysics and Research School
of Earth Sciences, The Australian National University, \\ 
Canberra ACT 0200, Australia \\
$^3$Department of Physics and Engineering, Macquarie University, Sydney
NSW 2109, Australia}

\date{2009 September 3}
\pagerange{\pageref{firstpage}--\pageref{lastpage}}
\pubyear{2009}
\begin{document}

\maketitle
\label{firstpage}
\begin{abstract}

We study a model of weakly ionized, protostellar accretion discs that
are threaded by a large-scale, ordered magnetic field and power a
centrifugally driven wind. We consider the limiting case where the wind
is the main repository of the excess disc angular momentum and
generalize the radially localized disc model of \citet{WK93}, which
focussed on the ambipolar diffusion regime, to other field diffusivity
regimes, notably Hall and Ohm. We present a general formulation of the
problem for nearly Keplerian, vertically isothermal discs using both the
conductivity-tensor and the multi-fluid approaches and simplify it to a
normalized system of ordinary differential equations in the vertical
space coordinate. We determine the relevant parameters of the problem
and investigate, using the vertical-hydrostatic-equilibrium
approximation and other simplifications, the parameter constraints on
physically viable solutions for discs in which the neutral particles are
dynamically well coupled to the field already at the midplane. When the
charged particles constitute a two-component ion--electron plasma one
can identify four distinct sub-regimes in the parameter domain where the
Hall diffusivity dominates and three sub-regimes in the Ohm-dominated
domain. Two of the Hall sub-regimes can be characterized as being
ambipolar diffusion-like and two as being Ohm-like: The properties of
one member of the first pair of sub-regimes are identical to those of
the ambipolar diffusion regime, whereas one member of the second pair
has the same characteristics as one of the Ohm sub-regimes. All the Hall
sub-regimes have $B_{r{\rm b}}/|B_{\phi{\rm b}}|$ (ratio of
radial-to-azimuthal magnetic field amplitudes at the disc surface) $>1$,
whereas in two Ohm sub-regimes this ratio is $< 1$.  When the
two-component plasma consists instead of positively and negatively
charged grains of equal mass, the entire Hall domain and one of the Ohm
sub-regimes with $B_{r{\rm b}}/|B_{\phi{\rm b}}|< 1$ disappear.  All
viable solutions require the midplane neutral--ion momentum exchange
time to be shorter than the local orbital time.  We also infer that
vertical magnetic squeezing always dominates over gravitational tidal
compression in this model.  In a follow-up paper we will present exact
solutions that test the results of this analysis in the Hall regime.

\end{abstract}

\section{Introduction}
\label{sec:intro}

Star formation is thought to be induced by the gravitational collapse of
the dense cores of molecular clouds \citep[e.g.][]{SAL87}. During this
process, angular momentum conservation results in the progressive
increase of the centrifugal force, a process that eventually halts the
infalling matter and leads to the development of a central mass
(protostar) surrounded by a rotationally supported, gaseous disc. In
the presence of an angular momentum transport mechanism, mass accretion
on to the protostar proceeds through this disc, and it is believed that
this is how stars typically gain most of their mass.

A common feature of accreting protostellar systems is their association
with energetic bipolar outflows that propagate along the rotation axis
of the source \citep[e.g.][]{BRD07}. The outflows from
low-bolometric-luminosity objects ($L_{\rm bol} < 10^3 L_{\odot}$) have
velocities in the range $\sim 150 - 400\;$ km s$^{-1}$ (of the order of
the escape speed from the vicinity of the central protostar) and are
typically well collimated (opening half-angles $\sim 3-5\,$degrees on
scales of $10^3 - 10^4\,$AU). The connection between accretion discs and
outflows, which extends smoothly from very low-mass stars and brown
dwarfs \citep[e.g.][]{MJB05} all the way up to protostars with masses of
at least $\sim 10\,M_\odot$ \citep[e.g.][]{CR98}, is manifested by the
observed correlations between accretion diagnostics (e.g.~inverse
P-Cygni profiles and excess emission in UV, IR and millimeter
wavelengths) and outflow signatures (e.g.~P-Cygni profiles, optical
forbidden lines, thermal radio radiation and the presence of molecular
lobes) in these objects \citep[e.g.][]{HEG95}. Also relevant are the
correlations of the type $\dot{M} \propto L_{\rm bol}^\delta$ (with
$\delta \sim 0.6 - 0.7$) that have been inferred for both accretion and
outflow rates in low- and high-$L_{\rm bol}$ objects
\citep[e.g.][]{Lev88} and the joint decline in outflow activity and disc
frequency (as well as in the inferred accretion rate) with stellar age
\citep[e.g.][]{CHS00,Cal05}.

These features point to an underlying physical link between accretion
and outflow processes in these systems. A natural explanation for this
connection is that outflows provide an efficient means of transporting
away the excess angular momentum and of tapping the liberated
gravitational potential energy of the accreted matter. In this scenario,
disc material is centrifugally accelerated by the torque exerted by a
large-scale, ordered magnetic field that threads the disc
(\citealt{BP82}, hereafter BP82; see also the reviews by \citealt{KP00},
\citealt{Pud07} and \citealt{KS09}). A dynamically significant, ordered
field component is indicated by the hourglass field morphology detected
by polarization measurements in several pre-collapse molecular cloud
cores on sub-parsec scales \citep[e.g.][]{Sch98,GRM06,Kir09}. These
interstellar field lines are expected to be advected inward when the
core undergoes gravitational collapse and could naturally give rise to
an ordered field in the resulting protostellar discs. Alternatively, an
ordered field component could be generated locally in the disc (and
star) by a dynamo process. The comparatively high momentum discharges
inferred in protostellar outflow sources are consistent with this
interpretation, as alternative wind driving mechanisms would fall far
short of the indicated requirements \citep[e.g.][]{KP00}. The
centrifugal wind picture can also naturally account for the
comparatively 
high ratio ($\sim 0.1$) of
the mass outflow to the mass accretion rate that is inferred in both the
quiescent and the outburst phases of low-mass protostars
\citep[e.g.][]{Ray07,HK96}.

Vertical angular-momentum transport by a large-scale, ordered magnetic
field is expected to dominate radial angular-momentum transport via
turbulence induced by the magnetorotational instability \citep[MRI;
e.g.][]{BH98} when the disc is threaded by a comparatively strong
magnetic field, corresponding to the ratio $a_0$ of the Alfv\'en
speed $v_{\rm A 0}$ (where the subscript 0 denotes the midplane) to the
isothermal sound speed $c_{\rm s}$ being not
much smaller than unity.\footnote{\label{fn:braking} Note that vertical
angular-momentum transport can also be mediated by additional mechanisms
(not considered in this paper), including magnetic braking (the
launching of torsional Alfv\'en waves into the ambient interstellar
medium; e.g. \citealt{KK02}), ``failed'' winds (which do not become
supersonic or super-Alfv\'enic) and non-steady phenomena.}  When this
condition is satisfied, angular momentum transport by a small-scale,
disordered magnetic field is suppressed because the wavelength of the
most unstable MRI perturbations exceeds the magnetically reduced disc
scale-height (\citealt[e.g.][]{WK93}, hereafter WK93). On the other
hand, when the magnetic field is comparatively weak ($a_0 \ll 1$),
MRI-induced radial transport should dominate.

Previous studies have considered the aforementioned radial and vertical
angular-momentum transport mechanisms under a number of simplifying
assumptions. The MRI has been analyzed in both its linear and non-linear
stages \citep[e.g.][]{SS02a,SS02b,SW03,SW05,FS03,San04,Des04,TSD07}.
Similarly, centrifugal wind-driving discs have been examined both
semi-analytically, making use of self-similarity formulations
\citep[e.g.][]{WK93,Li95,Li96,Fer97}, and through numerical simulations
that utilize resistive-MHD codes
\citep[e.g.][]{CK02,Kuw05,MCS06,Zan07}. It is, however, likely that both
the radial and vertical modes of angular momentum transport play a role
in real accretion discs. While quasi-steady disc models in which both
mechanisms operate had been considered in the literature
\citep[e.g.][]{LRN94,CF00,OL01}, they did not specifically link the
radial transport of angular momentum with its origin in MRI-induced
turbulence. In a first attempt in this direction, \citet{SKW07} explored
the possibility that radial and vertical transport could operate at the
same radial location in a protostellar disc. They derived an approximate
criterion for identifying the vertical extent of a wind-driving disc at
the given radius that is unstable to MRI-induced turbulence and obtained
a quantitative estimate (based on the numerical simulations of
\citealt{San04}) of the amount of angular momentum that is removed
radially from that region.

An important consideration for the analysis of magnetic angular momentum
transport in protostellar discs is that these discs are generally weakly
ionized over most of their extents, so magnetic diffusivity effects are
significant. This is a critical issue because both the centrifugal-wind
and MRI-turbulence transport mechanisms require a minimum level of
field--matter coupling to be effective. A complete model therefore needs
to take into account the detailed ionization and conductivity structure
of the disc/wind system. At low densities (near the disc surfaces and at
large radii) the disc is typically in the {\em ambipolar diffusion}
regime. As the density and the shielding column increase (closer to the
disc midplane and the central source), the fluid passes successively
through the {\em Hall} and {\em Ohm} diffusivity regimes (see
Section~\ref{sec:formulation}). Note, however, that the disc parameters
may well be such that the gas does not clearly correspond to any one of
these limiting cases and must instead be described more generally using
either the conductivity-tensor (Section~\ref{subsec:model}) or the
multi-fluid-decomposition (Section~\ref{subsec:multi}) formulations. Most
previous treatments have simplified the problem by focussing on a single
conductivity regime. In particular, WK93 and \citet{SKW07} concentrated
on the ambipolar diffusion limit (although WK93 also discussed
Hall-current effects). WK93 were able to obtain simple analytic
constraints on the parameter values for physically viable wind-driving disc
solutions in this limit by modelling the vertical structure of the disc
under the {\em hydrostatic} (negligible vertical velocity)
approximation. One of their derived constraints yielded the minimum
required degree of field--matter coupling, and they proceeded to verify
their analysis by obtaining numerical solutions for discs in this {\em
strong coupling} regime.

Although it is in principle possible to seek a self-similar solution for
the global disc/wind configuration \citep[e.g.][]{Kon89,Fer97},
WK93 adopted a somewhat simpler approach in which they obtained a
solution for the vertical structure of a radially localized (radial
extent $\Delta r \ll r$ at a cylindrical radius $r$) region of the disc
and matched it to a self-similar (in the spherical radial coordinate
$R$) global wind solution of the type originally constructed by
BP82. This approach facilitated the derivation of viability and
consistency constraints involving exclusively the disc model parameters,
and subsequently \citet{Li96} demonstrated that the basic features of this
solution are consistent with those of a fully global disc/wind
similarity solution.  Our aim in this paper is to generalize the
parameter constraints derived in WK93 for the ambipolar diffusion regime
to the other basic diffusivity regimes, namely Hall and Ohm. In each of
the latter regimes we identify parameter combinations that yield
different sets of constraints, which correspond to solutions with
distinct physical properties, and we use these constraints to delineate
the Hall and Ohm sub-regimes in parameter space where one can derive
viable, strongly coupled, wind-driving disc solutions. In a follow-up
paper (\citealt{SWK09}, hereafter Paper~II) we test the results for the
Hall regime -- derived by using the hydrostatic approximation -- against
full numerical solutions obtained in the radially localized disc
approximation of WK93.

As noted above, the Hall and Ohm diffusivity regimes are likely to apply
near the midplane in the inner regions of real protostellar
discs. Typically, the surface regions from which the wind is driven
would be sufficiently well ionized by cosmic rays and by the
protostellar radiation field (see Section~\ref{subsec:model}) to lie in
the ambipolar diffusion regime \citep[e.g.][]{SW05,KS09}. Nevertheless,
we adopt the approximation that the entire disc cross section between
the midplane and the base of the wind lies in a single diffusivity
sub-regime in order to bring out the distinct properties of the
solutions for each parameter range.  This approximation is less
appropriate for the Ohm regime, both because {\em two} additional
regimes (ambipolar diffusion {\em and} Hall) would be encountered on the
way up to the disc surface in this case and because the column densities
of wind-driving discs are relatively small (especially in comparison
with those typically implied by radial-turbulent-transport models) and
are therefore likely to contain only limited regions (if any at all)
where the Ohm diffusivity dominates. We have nevertheless chosen to
include the Ohm regime in our analysis of weakly ionized discs for
completeness as well as for possible comparisons with resistive-MHD
numerical simulations.\footnote{Note, however, that we do not continue
to consider this case in Paper~II, which concentrates on the
ambipolar-diffusion and Hall regimes.} The results of this analysis
could potentially be applicable also to the collisionally ionized
innermost region of the disc, where anomalous resistivity (the enhanced
drag between ions and electrons due to scattering off electromagnetic
waves generated by current-driven plasma instabilities) might
develop. To obtain explicit solutions in the different diffusivity
regimes, we have generalized the WK93 model setup by implementing a
conductivity-tensor scheme that can be applied to any given vertical
ionization structure. In Paper~II we use a simplified version of this
scheme to mimic the single-diffusivity conditions adopted in the present
work.

This paper is organized as follows. Section~\ref{sec:formulation}
summarizes the formulation, including the disc model and the governing
equations. Section~\ref{sec:method} describes the methodology we employ
to reduce these equations to a set of ordinary differential equations in
the vertical coordinate $z$ and to express them in a dimensionless form;
it also lists the parameters of the
problem. Section~\ref{sec:constraints} derives the constraints that
characterize the parameter space where physically viable solutions exist and
explicates them for the Hall regime, with the corresponding constraints
for the Ohm regime presented in Section~\ref{sec:Ohm_constraints}. The key
findings of the paper are discussed in section~\ref{sec:discuss} and
summarized in Section~\ref{sec:conclude}.

\section{Formulation}
\label{sec:formulation}

\subsection{Disc model}
\label{subsec:model}

We model the disc as being in a steady state, geometrically thin,
vertically isothermal, nearly Keplerian and in dynamical equilibrium in
the gravitational potential of the central protostar.  Since we expect
all the physical variables to exhibit smooth spatial variations, the
assumption of geometrical thinness enables us to neglect the radial
derivative terms ($|\partial/\partial r|\sim 1/r$) in our equations in
comparison with the vertical derivative terms ($|\partial/\partial z|
\sim 1/h$), where $r$ and $z$ are cylindrical coordinates and $h(r)$
($\ll r$) is the disc pressure (or density) scale-height at the radius
$r$.\footnote{The only exception to this approach is our retention of
the radial derivative of the azimuthal velocity component $\vf$ (see
WK93).} The disc material is assumed to be weakly ionized, with the
abundances of charged species being so low that (i) the effect of
ionization and recombination processes on the neutral gas can be
neglected and (ii) the inertial, gravitational and thermal forces on the
ionized species are negligible in comparison with the electromagnetic
force exerted by a large-scale, ordered magnetic field. Under these
approximations, separate equations of motion for the charged species are
not required, and their effect on the neutrals can be incorporated via a
conductivity tensor $\bmath{\sigma}$ (e.g. \citealt{Cow76};
\citealt{War99} and references therein), which is a function of position
$\{r,z\}$. This formulation makes it possible to systematically
incorporate different ionized fluid components as well as the three
basic field--matter diffusion mechanisms (ambipolar, Hall and Ohm). The
conductivity tensor components are the Ohm, Hall and Pedersen terms,
given by
\begin{equation}
\sigma_{\rm O} = \frac{ec}{B}\sum_{j} n_j |Z_j| \beta_j \, ,
\label{eq1:sigma0}
\end{equation}
\begin{equation}
	\sigma_{\rm H} = \frac{ec}{B}\sum_{j}\frac{n_j Z_j}{1+\beta_j^2}\, 
	\label{eq1:sigma1}
\end{equation}
and
\begin{equation}
	\sigma_{\rm P} = \frac{ec}{B}\sum_{j}\frac{n_j |Z_j|
\beta_j}{1+\beta_j^2} \, ,
	\label{eq1:sigma2}
\end{equation}
respectively \citep[e.g.][]{WN99}. In the above
expressions, $n_j$ is the number density of charged species $j$ (of
mass $m_j$ and total electric charge $Z_j e$), $c$ is the speed of
light, and  
\begin{equation}
\beta_j= \frac{|Z_j| eB}{m_j c} \, \frac{1}{\gamma_j \rho}
\label{eq:Hall_parameter}
\end{equation}
is the Hall parameter, the ratio of the gyrofrequency and the
collision frequency of species $j$ with the neutrals, which
measures the relative importance of the Lorentz and drag
forces on the motion of the charged species. In these equations 
\begin{equation}
B \equiv |\mathbf{B}|\, {\rm sgn}\{B_z\}
\label{eq:B}
\end{equation}
is the {\em signed} magnetic field amplitude, with the sign
introduced to keep the dependence of the Hall conductivity on the
magnetic field polarity.\footnote{Note from
equations~(\ref{eq1:sigma0})--(\ref{eq1:sigma2}) that, in contrast with
$\sigma_{\rm H}$, which depends on an odd power of $B$, both the Ohm and
Pedersen conductivities are independent of the magnetic field polarity.}
Also, in equation (\ref{eq:Hall_parameter}),
\begin{equation}
\gamma_j = \frac{<\sigma v>_j}{m_j+m} \ ,
	\label{eq:gamma}
\end{equation}
where $m = 2.33\, m_{\rm H}$ is the mean mass of the neutral particles
in terms of the hydrogen nucleus mass $m_{\rm H}$, $\rho$ is the density
of the neutral gas and $<\sigma v>_j$ is the rate coefficient of
momentum exchange of species $j$ with the neutrals.

The above equations can be used to treat charged species with a wide
range of masses and degrees of coupling with the neutral gas. However,
to simplify the analysis and make it more transparent, we henceforth
specialize to a fluid with only two charged species, one with positive
charge (which we refer to as ``ions'' and denote by a subscript i;
$Z_{\rm i} > 0$) and the other with negative charge (``electrons,''
subscript e, satisfying $Z_{\rm e} < 0$). This description is a good
approximation both in the comparatively low-density regions of
protostellar discs, where the two species can be taken as metal ions (of
typical mass $m_{\rm i} = 30 \, m_{\rm H}$; e.g. \citealt{DRD83}) and
electrons, and in high-density ($\rho \gtrsim 10^{-13}\, {\rm g\,
cm}^{-3}$) regions where the dominant charge carriers are positively and
negatively charged dust grains of equal mass
\citep[e.g.][]{UN90,NH94,KK02}.\footnote{\label{fn:dust} During the
late ($\gtrsim 10^5 \,$yr) phases of protostellar evolution, after the
dust has largely settled to the midplane
\citep[e.g.][]{NNH81,DD04,Dal06}, ions and electrons could dominate the
charged species also in the dense inner regions of the disc.}  The
``ion'' and ``electron'' Hall parameters are related by $\beta_{\rm i} =
q \beta_{\rm e}$, where
\begin{equation}
q = \frac{m_{\rm e}}{m_{\rm i}}\frac{m_{\rm i} + m}{m_{\rm e} +
m}\frac{<\sigma v>_{\rm e}}{<\sigma v>_{\rm i}} \, .
\label{eq:q}
\end{equation}
In the case of charged grains of equal and opposite charge as well as equal
mass, $q=1$, whereas in the case of ions and electrons we find, using
\begin{equation}
<\sigma v>_{\rm i} = 1.6\ee {-9} \ut cm 3 \ut s -1
	\label{eq:sigvi} 
\end{equation}
and
\begin{equation}
	<\sigma v>_{\rm e} \approx (1\ee {-15} \ut cm 2 )
	\left(\frac{128 kT}{9\pi m_{\rm e}}\right)^{1/2} 
	\label{eq:sigve}
\end{equation}
from the results of \citet{DRD83}, that $q \approx 1.3\ee{-4}
\sqrt{T}$ (where $T$ denotes the gas temperature and $k$ is
Boltzmann's constant). In our analysis we often encounter the product
$s \equiv \beta_{\rm e} \beta_{\rm i} = \beta_{\rm i}^2/q$, which is
sometimes referred to in the literature as the {\em ion slip factor}
\citep[e.g.][]{MK73}.

With the help of the charge-neutrality condition, 
\begin{equation}
\sum_j n_j Z_j = 0 \,, 
\label{eq:neutrality} 
\end{equation} 
equations~(\ref{eq1:sigma0})--(\ref{eq1:sigma2}) become 
\begin{equation}
\sigma_{\rm O} = \frac{cen_{\rm e} |Z_{\rm e}|}{B} (\beta_{\rm e} +
\beta_{\rm i}) \, , 
\label{eq:sig1} 
\end{equation} 
\begin{equation} 
\sigma_{\rm H} = \frac{cen_{\rm e}|Z_{\rm e}|}{B} \frac{(\beta_{\rm e} +
\beta_{\rm i})(\beta_{\rm e} - \beta_{\rm i})}{(1+\beta_{\rm
e}^2)(1+\beta_{\rm i}^2)}
\label{eq:sig2} 
\end{equation} 
and 
\begin{equation}
\sigma_{\rm P} = \frac{cen_{\rm e}|Z_{\rm e}|}{B} \frac{(1+\beta_{\rm
i}\beta_{\rm e})(\beta_{\rm e} + \beta_{\rm i})}{(1+\beta_{\rm
e}^2)(1+\beta_{\rm i}^2)}\, , 
\label{eq:sig3} 
\end{equation} 
respectively \citep[e.g.][]{SW03}. We also make use of
$\sigma_{\perp} = \sqrt{\sigma_{\rm H}^2 + \sigma_{\rm P}^2}$, the total
conductivity perpendicular to the magnetic field, which is given by
\begin{equation} 
\sigma_\perp = \frac{cen_{\rm e}|Z_{\rm e}|}{B} \frac{(\beta_{\rm e} +
\beta_{\rm i})}{[(1+\beta_{\rm e}^2)(1+\beta_{\rm i}^2)]^{1/2}} \; .
\label{eq1:sigperp} 
\end{equation} 
The ratios of the conductivity tensor components $\sigma_{\rm H}$,
$\sigma_{\rm P}$ and $\sigma_{\rm O}$ determine the relative importance
of the ambipolar, Hall and Ohm diffusion mechanisms
\citep[e.g.][]{WN99}.  These ratios, in turn, depend on the magnitudes
of the ``ion'' and ``electron'' Hall parameters, which are functions of
the neutral density and magnetic field amplitude. The spatial variation
of the latter quantities in protostellar systems results in different
diffusion mechanisms dominating in different regions of the disc. We now
briefly outline the main properties of these different regimes, although
it must be borne in mind that some regions of the disc will inevitably
correspond to the transition zones between two such regimes (or between
any of the sub-regimes that we identify) and hence cannot be so simply
classified.

\begin{enumerate}
 
\item \emph{Ambipolar diffusion}. This regime dominates when
$\sigma_{\rm O} \gg \sigma_{\rm P} \gg |\sigma_{\rm H}|$ or,
equivalently, when $|\beta_{\rm i}| \gg 1$.  In this case the magnetic
field is effectively frozen into the ionized fluid component (the
``electrons'' and the ``ions'') and drifts with it through the
neutrals. This regime is expected to dominate at relatively low
densities in protostellar discs -- throughout the vertical column in the
outermost regions of the disc (e.g. WK93) and close to the disc surfaces
at smaller radii.

\item \emph{Hall diffusion} ($\sigma_{\rm P} \ll |\sigma_{\rm H}| \ll
\sigma_{\rm O}$).  These conditions are satisfied when the most highly
mobile charged particles (having the highest value of $|Z_je/m_j|$ --
the electrons in the case of an ion--electron plasma) are well coupled
to the magnetic field ($|\beta_{\rm e}| \gg 1$) even as the drift of the
more massive charged particles of the opposite sign (i.e. the ions in an
ion--electron plasma) through the neutrals is strongly inhibited by
collisions ($|\beta_{\rm i}| \ll 1$). In this limit the current in an
ion--electron plasma is dominated by the electrons, which drift
perpendicularly to the magnetic and electric fields (thereby minimizing
the electromagnetic force acting on them, with the electron--neutral
drag force typically remaining negligible). This gives rise to the Hall
current, which is perpendicular to both the electric and magnetic
fields. This regime typically dominates close to the disc midplane on
``intermediate'' radial scales
\citep[e.g.][]{Li96,War99,SS02a,SW05}. Note that $\sigma_{\rm H}$ (and
therefore the Hall diffusivity regime) vanishes identically when
$|\beta_{\rm i}| = |\beta_{\rm e}|$ (i.e. when $q=1$).

\item \emph{Ohm diffusion} ($\sigma_{\rm O} \approx \sigma_{\rm P} \gg
|\sigma_{\rm H}|$ or, equivalently, $|\beta_{\rm e}| \ll 1$). In this
case all the ionized species are uncoupled from the field by collisions
with the neutrals, so the magnetic field cannot be regarded as being
frozen into any of the fluid components. In this parameter regime one
can formulate the problem in terms of a simple scalar conductivity,
corresponding to the familiar ohmic resistivity. However, as we show in
Section~\ref{sec:Ohm_constraints}, one cannot in general neglect the
Hall term in Ohm's law in classifying the physically viable disc
solutions in this regime. The Ohm diffusivity may dominate in the most
shielded inner regions of particularly massive protostellar discs (but
outside the radius where collisional ionization of alkali metals sets
in). As we noted in Section~\ref{sec:intro}, it is also conceivable that
{\em anomalous} Ohm diffusivity dominates in the collisionally ionized
zone further in.  In fact, certain models \citep[e.g.][]{BL94,Gam99}
link the formation of a higher ionization zone in the innermost region
of the disc to the development of FU Orionis-type outbursts, during
which most of the accretion on to the central protostar may take place
\citep[e.g.][]{CHS00}.

\end{enumerate} 

Note that the ion slip factor $s = \beta_{\rm i}^2/q$ is $\gg 1$ in the
ambipolar diffusion regime and $\ll 1$ in the Ohm regime. This factor
can be either $< 1$ or $> 1$ for a $q \ll 1$ two-component plasma in the Hall
regime. This suggests that in the latter case one can distinguish
between the ambipolar diffusion-modified Hall regime, in which the
inequalities $\sqrt{q} \ll |\beta_{\rm i}| \ll 1$ are satisfied, and the
Ohm diffusion-modified Hall regime, in which the inequalities $q \ll
|\beta_{\rm i}| \ll \sqrt{q}$ are obeyed \citep{Kon97}. We return to
provide explicit support for this characterization in
Section~\ref{sec:discuss}.

In the fiducial minimum-mass solar nebula model \citep{Hay81} the
ambipolar, Hall, and Ohm diffusivity regimes dominate at the disc
midplane on scales $r \gtrsim 10\,$AU, $\sim 1 - 10\,$AU and $\sim 0.1 -
1\,$AU, respectivley, for likely magnetic field strengths and when the
effect of grains is neglected. If sufficiently small ($\lesssim
0.1\,\mu$m) grains are present at $z=0$ and the density is large enough
that they carry a significant fraction of both the negative {\em and}
the positive charges then the extent of the midplane Hall regime is
strongly reduced \citep[e.g.][]{WN99,War07,SW08}.  Wind-driving
accretion-disc models typically have lower column densities and midplane
densities than either the minimum-mass model or standard viscous disc
models (a reflection of their comparatively high inward drift speeds;
see Section~\ref{sec:discuss}): this has the effect of pushing both the
inner and outer edges of the midplane Hall regime inward. In practice,
the spatial distribution of the different diffusivity regimes depends on
the mass accretion rate, the dust content and the nature of the
ionization mechanisms that affect the disc, and it could vary from
source to source and during the evolution of any particular system.

Evaluating the conductivity tensor in a real disc involves calculating
the ionization balance and associated abundances of charged species at
any given location. This balance is, in turn, the result of a delicate
equilibrium between ionization and recombination processes acting both
in the gas phase and on the surfaces of dust grains (if present). The
most relevant ionization sources outside the central $\sim 0.1\,$AU of
the disc are nonthermal: X-rays and UV radiation, emanating from the
magnetically active young star and its magnetosphere, as well as
interstellar cosmic rays and radioactive decay inside the disc
\citep[e.g.][]{Hay81, UN81,UN90,IG99}. In the vicinity of the star the
disc surface is ionized mainly by X-rays, but UV radiation also becomes
important at larger distances \citep[e.g.][]{SWH04,ACP05,Gla05}. Within
the innermost $\sim 0.1\,$AU the degree of ionization could be
significantly raised by collisional ionization of alkali metals,
particularly potassium \citep[e.g.][]{Gam96,Li96}. A detailed
calculation of the vertical ionization and conductivity structure of a
wind-driving disc has so far been performed only in the context of a
radially localized treatment and without including dust
\citep[e.g.][]{KS09}; a generalization to a global disc/wind model and
the incorporation of dust remain to be done.

In this paper and in Paper~II we adopt a simpler treatment than the one
outlined above, motivated by the fact that our main interest is to (i)
identify the regions in parameter space where viable wind-driving disc
solutions can be found under weak-ionization conditions and (ii) derive
the main properties of these solutions. As our classification scheme
requires the specification, at the midplane of the disc, of the values
of three distinct variables (see Section~\ref{sec:discuss}), we make the
assumption that the density and ionization structure of the disc is such
that, at the radial location under consideration, the parameter regime
determined by the midplane values of the chosen three variables
continues to apply throughout the vertical extent of the disc. In the
analysis presented in Sections~\ref{sec:constraints}
and~\ref{sec:Ohm_constraints} of this paper we use the variables
$\beta_{\rm e}$, $\beta_{\rm i}$ and $\Upsilon$, the neutral--ion
coupling parameter defined in equation~(\ref{eq:eta}), and we assume
that they remain constant with height at least within the
quasi-hydrostatic layer. In the numerical work presented in Paper~II we
choose the distinct variables to be two independent ratios of the
conductivity tensor components and the field--matter coupling parameter
$\Lambda$ (the Elsasser number; see Section~\ref{subsec:param}), and we
assume that they remain constant between $z=0$ and the sonic point above
the disc surface. Notwithstanding the difference in the specific
variables chosen in each case, these two procedures are effectively
equivalent.

\subsection{Governing Equations}
\label{subsec:governing}

Our basic equations describe the conservation of mass
\begin{equation}
\delt{\rho} + \div(\rho \v) = 0 
	\label{eq:continuity}
\end{equation}
and momentum
\begin{equation}
\delt{\v} + (\v \bdot \grad)\v + \frac{1}{\rho}\grad P
	+\grad \Phi - \frac{\J\cross\B}{c\rho} = 0
	\label{eq:momentum}
\end{equation}
for the neutral gas, as well as the evolution of the magnetic field
\begin{equation}
\delt{\B} = \curl \E = \curl (\v \cross \B) - c \curl \E' \ ,
	\label{eq:induction}
\end{equation}
where $v$ is the fluid velocity and $\J$ is the current density. 
In the equation of motion (equation~\ref{eq:momentum})
\begin{equation}
P \equiv \rho c_{\rm s}^2 = \frac{\rho k T}{\mu m_{\rm H}}
\end{equation}
is the gas pressure, $c_{\rm s}$ is the isothermal sound speed and $\mu
\equiv m/m_{\rm H} = 2.33$ is the molecular weight.
In addition, $\Phi$ is the gravitational 
potential of the central object, given by
\begin{equation}
\Phi = -\frac{GM}{(r^2 + z^2)^{1/2}} \ ,
          \label{eq:gravpotential}
\end{equation}
where $G$ is the gravitational constant and $M$ is the mass of the
protostar (which is treated as a point mass at the origin of the
coordinate system).  Note that in equation~(\ref{eq:momentum}) we have
made use of the low-inertia limit for the ionized species by balancing
the sum of the ion--neutral and electron--neutral drag forces
($\bmath{f}_{\rm in}$ and $\bmath{f}_{\rm en}$, respectively) by the
Lorentz force,
\begin{equation}
\bmath{f}_{\rm in} + \bmath{f}_{\rm en} = \frac{\J \cross \B}{c} \ ,
\label{eq:lorentz}
\end{equation}
on the assumption that all other terms in the momentum equations for the
charged particles remain small.

In the induction equation (\ref{eq:induction})
\begin{equation}
\E' = \E + \frac{\v \cross \B}{c}
	\label{eq:E}
\end{equation}
is the electric field in the frame comoving with the neutrals and $\E$
is the corresponding field in the inertial (laboratory) frame. The field
$\E'$ is related to the current density $\J$ through the conductivity tensor
$\bmath{\sigma}$ (see  equation~\ref{eq:J-E} below). Furthermore,
the magnetic field satisfies the solenoidal condition
\begin{equation}
\div \B = 0 \, ,
\label{eq:divB}
\end{equation}
which implies (in analogy with the inference on $\rho \vz$ following
equation~\ref{eq:continuity_simp} below) 
that in a thin disc $B_z$ is approximately constant with
height, whereas the current density obeys Amp\`ere's law
\begin{equation}
\J = \frac{c}{4\pi}\grad\cross \B
\label{eq:j_curlB}
\end{equation}
(where we neglected the displacement current) and Ohm's law,
\begin{equation}
	\J = \bmath{\sigma}\bdot \E' = \sigma_{\rm O} \Epa +
	\sigma_{\rm H} \Bh \cross \Epe + \sigma_{\rm P} \Epe  \, .
	\label{eq:J-E}
\end{equation}

Equation (\ref{eq:J-E}) can be inverted to yield the following 
expression for $\E'$:
\begin{equation}
	\E' = \frac{\J}{\sigma_{\rm O}} +
	      \frac{\sigma_{\rm H}}{\sigma_\perp^2} \frac{\J\cross\B}{B} -
	\left(\frac{\sigma_{\rm P}}{\sigma_\perp^2}-
	\frac{1}{\sigma_{\rm O}}\right)
	\frac{(\J\cross\B)\cross\B}{B^2} \ .
	\label{eq1:E-J}
\end{equation}
The terms on the right-hand side of equation~(\ref{eq1:E-J}) are, from
left to right, the resistive (Ohm), Hall and ambipolar diffusion
contributions, respectively. 
Substituting equations~(\ref{eq:sig1})--(\ref{eq1:sigperp}) for a
two-component plasma into
equation~(\ref{eq1:E-J}) gives
\begin{equation}
\E' = \frac{\beta_{\rm e}}{\beta_{\rm e} + \beta_{\rm
i}}\,\frac{\J}{\sigma_{\rm e}} + \frac{\beta_{\rm e}-\beta_{\rm i}}{\beta_{\rm
e}+\beta_{\rm i}} \, \frac{\J\cross\B}{c e n_{\rm e} |Z_{\rm e}|} - 
\frac{\beta_{\rm e} \beta_{\rm i}}{\beta_{\rm e} +
\beta_{\rm i}}\,\frac{(\J\cross\B)\cross\B}{B c e n_{\rm e} |Z_{\rm e}|}\ ,
	\label{eq1:E-J2}
\end{equation}
where $\sigma_{\rm e}$ is the ``electron'' electrical conductivity,
\begin{equation}
\sigma_{\rm e} = \frac{n_{\rm e} e^2 Z_{\rm e}^2}{m_{\rm e} \gamma_{\rm
e} \rho} \ .
\label{eq:sigma_e}
\end{equation}
Using $\beta_{\rm i}/(Bce n_{\rm e} |Z_{\rm e}|) = (c^2 \gamma_{\rm i}
\rho \rho_{\rm i})^{-1}$ in the last term of equation~(\ref{eq1:E-J2})
and taking the limit $q \ll 1$ leads to 
\begin{equation}
\E' = \frac{\J}{\sigma_{\rm e}} +
\frac{\J\cross\B}{c e n_{\rm e} |Z_{\rm e}|} -
	       \frac{(\J\cross\B)\cross\B}{c^2 \gamma_{\rm i} \rho_{\rm
	       i} \rho} \ ,
	\label{eq1:E-J3}
\end{equation}
which is similar in form to expressions obtained in multi-fluid
formulations \citep[e.g.][]{Kon89,BT01}. 

The impact of the magnetic diffusivity on the evolution of the fluid can
be more readily appreciated if we use the Ohm, Hall and ambipolar
\emph{diffusivity} terms, given by
\begin{equation}
\eta_{\rm O} = \frac{c^2}{4 \pi \sigma_{\rm O}} \; ,
\label{eq:etaO}
\end{equation}
\begin{equation}
\eta_{\rm H} = \frac{c^2}{4 \pi \sigma_{\perp}} \frac{\sigma_{\rm
H}}{\sigma_{\perp}}
\label{eq:etaH}
\end{equation}
and
\begin{equation}
\eta_{\rm A} = \frac{c^2}{4 \pi \sigma_{\perp}} \frac{\sigma_{\rm
P}}{\sigma_{\perp}} -  \eta_{\rm O} \, .
\label{eq:etaA}
\end{equation}
Expressing the conductivity terms in equation (\ref{eq1:E-J}) as
functions of the Ohm, Hall and ambipolar diffusivities
(equations~\ref{eq:etaO}--\ref{eq:etaA}) and substituting the resulting
expression for $\E'$ into equation~(\ref{eq:induction}) leads  
to the following form of the induction equation \citep{War07}:
\begin{eqnarray}
\lefteqn{\delt{\B} = \curl \left (\v \cross \B\right) - \curl \left[
\eta_{\rm O} \curl \B \right. } 
	\nonumber\\
& & {} + \left.  \eta_{\rm H} (\curl \B) \cross \hat{\B} + \eta_{\rm A}
(\curl \B)_{\perp} \right] \; ,
	\label{eq1:induction_drift}
\end{eqnarray}
where the subscript $\perp$ again denotes the direction perpendicular to
$\B$. In the limit of a two-component plasma with $q\ll 1$ it can be shown
\citep{War07} that
\begin{equation}
\etaH = \beta_{\rm e} \etaO
\label{eq:etaH1}
\end{equation}
and 
\begin{equation}
\etaA =  \beta_{\rm e} \beta_{\rm i} \etaO \,.
\label{eq:etaA1}
\end{equation}

Under ideal-MHD conditions all the magnetic diffusivity terms in the induction
equation tend to zero and the evolution of the magnetic field is fully
determined by the inductive term (the first term on the right-hand side of
equations~\ref{eq:induction} and~\ref{eq1:induction_drift}).

\section{Methodology}
\label{sec:method}
\subsection{Simplification of the equations}
\label{subsec:ODE}

We now proceed to reduce the system of equations (\ref{eq:continuity})
-- (\ref{eq:induction}) and (\ref{eq:j_curlB}) to a set of ordinary
differential equations (ODEs) in $z$.  We closely follow the approach of
WK93 and consider, as in that paper, a radially localized region of the
disc that is threaded by an open magnetic field with an ``even''
symmetry ($B_{r 0}=B_{\phi 0}=0$).  Sections \ref{subsec:cont} --
\ref{subsec:induc_phi} deal with the mass and momentum conservation
equations for the neutral gas and with the induction equation that
describes the evolution of the magnetic field. The remaining two
subsections (\ref{subsec:amp} and \ref{subsec:ohm}) are concerned with
Amp\`ere's and Ohm's laws, respectively.

\subsection{Continuity}
\label{subsec:cont}

We are looking for steady-state ($\partial/\partial t = 0$),
axisymmetric ($\partial/\partial \phi = 0$) solutions. In this limit the
continuity equation (\ref{eq:continuity}) reads
\begin{equation}
\frac{1}{r}\frac{\partial}{\partial r} (r \rho v_r) + 
\frac{\partial}{\partial z} (\rho v_z) = 0 \, .
\label{eq:continuity_simp}
\end{equation}
This expression is further simplified by neglecting the radial
derivative term (see Section~\ref{subsec:model}), which implies that
$\rho v_z$ is constant with height. This approximation, first adopted in
WK93, has been critiqued \citep[e.g.][]{Fer97} for not allowing $v_z$ to
assume negative values within the disc, as it must do in cases (expected
to be typical) in which the disc thickness decreases as the protostar is
approached.  This issue can be fully addressed only in the context of a
global disc/wind model. However, it can be expected that any error
introduced by this approximation would be minimized if the upward mass
flux remained small enough for $v_z$ to have only a weak effect on the
behaviour of the other variables within the disc. Under the assumption
that $|\vr|$ is of the same order of magnitude as $|\vf - \vk|$ one can
readily show that the condition for this to hold is that $\vz/c_{\rm s}$
remain $\ll 1$ everywhere within the disc. This can be checked a
posteriori for each derived solution of the radially localized model,
and we do this for the solutions presented in Paper~II. Note that the
fulfillment of this condition also implies that the hydrostatic
simplification employed in the derivation of the parameter constraints
in Sections~\ref{sec:constraints} and~\ref{sec:Ohm_constraints} should
result in a good approximation to the complete numerical solution for
the vertical structure of the disc.

\subsection{Radial component of the momentum equation}

\begin{equation}
\rho v_r \delr{v_r} + \rho v_z \delz{v_r} - \frac{\rho v_{\phi}^2}{r} + \delr{P} 
+ \rho \delr{\Phi} = 
\frac{1}{c}( J_{\phi} B_z - J_z B_{\phi}) \, .
\end{equation}
The dominant terms in this equation are the gravitational and
centrifugal forces. We can thus simplify it by neglecting the $\partial
P/\partial r$ and $\partial v_r/\partial r$ terms. Because of the
assumed geometrical thinness, we can also neglect a term of order
$(z/r)^2 \vk^2/r$ in $\partial\Phi/\partial r$. However, to handle the
departure from exact Keplerian rotation, we keep the drag force
(i.e. the Lorentz force, communicated to the neutrals by ion--neutral
collisions; see equation \ref{eq:lorentz}) and the $\partial
v_r/\partial z$ term. Introducing these changes and using $\vk^2 -
v_{\phi}^2 \approx 2 \vk(\vk-v_{\phi})$, the radial momentum equation becomes
\begin{equation}
\rho v_z \frac{d v_r}{dz} +  \frac{2 \rho \vk}{r} (\vk - v_{\phi}) = 
\frac{1}{c}( J_{\phi} B_z - J_z B_{\phi}) \, .
\end{equation}

\subsection{Azimuthal component of the momentum equation}
\label{subsec:mom_phi}

\begin{equation}
\rho v_r \delr{\vf} + \frac{\rho \vf \vr}{r} + \rho \vz \delz{\vf} = 
\frac{1}{c}(J_z B_r - J_r B_z) \, .
\label{eq:azimom}
\end{equation}
In this equation, the torque due to the ion-neutral drag (the right-hand
side) and the inward radial transport are important in the hydrostatic
layer, whereas the vertical material transport becomes relevant as the
wind region is approached. It is thus appropriate to retain all the
terms.  Using $\vf \approx \vk$, we estimate $\partial(r \vf)/\partial r
\approx \vk/2$. With these changes, equation (\ref{eq:azimom}) reads
\begin{equation}
\frac{\rho \vr \vk}{2r} + \rho \vz \frac{d \vf}{dz} = 
\frac{1}{c}(J_z B_r - J_r B_z) \, .
\end{equation}

\subsection{Vertical component of the momentum equation}

\begin{equation}
\rho \vr \delr{\vz} + \rho \vz \delz{\vz} + \delz{P} + \rho \delz{\Phi} = 
\frac{1}{c}(J_r B_{\phi} - J_{\phi} B_r) \, .
\end{equation}
The dominant terms in this equation are the thermal pressure gradient
and the tidal and magnetic compression terms. This equation can be
simplified using

\begin{displaymath}
\delz{P} = c_{\rm s}^2 \delz{\rho}\ , \qquad
\delr{\vz} \approx 0\ ,
\end{displaymath}

\begin{displaymath}
\rho \vz \delz{\vz} \approx - \vz^2\delz{\rho}\ , \qquad
\rho \delz{\Phi} \approx \frac{\rho \vk^2}{r}\frac{z}{r} \ ,
\end{displaymath}
where the approximation  $\rho \vz = \text{const}$ was used to obtain
the first expression on the second line. The simplified equation is then
\begin{equation}
(c_{\rm s}^2 - v_z^2)\frac{d\rho}{dz} + \frac{\rho \vk^2}{r}\frac{z}{r} = 
\frac{1}{c}(J_r B_{\phi} - J_{\phi} B_r) \, .
\end{equation}

\subsection{Vertical component of the induction equation}
\label{subsec:induc_z}
\begin{equation}
 \delt{B_z} = - \frac{1}{r}\frac{\partial}{\partial r}(rcE_{\phi})\, ,
\label{eq:ind_z}
\end{equation}
where 
\begin{equation}
c E_\phi = \vr B_z -  \vz B_r + c E_{\phi}^\prime\, .
\label{eq:E_phi}
\end{equation}
In strictly steady-state discs $\grad \cross \E = - (1/c) \partial
\mathbf{B}/\partial t = 0$, so $\E$ can be 
expressed as a gradient of a scalar function and hence, under the
assumption of axisymmetry, $E_\phi = 0$. However, $B_z$ or,
equivalently, the poloidal (subscript p) magnetic flux function $\Psi(r,z)$
(defined by $2\pi \Psi = \iint \B_{\rm p} \bdot d{\bmath{S}}$, where the
integral is over a spatially fixed surface that is threaded by the
poloidal magnetic field) may still change on the
``long'' accretion time-scale $\tau_{\rm a} \equiv r/|v_r|$, corresponding to
a radial drift of the poloidal magnetic field through the disc. The
midplane radial speed of the magnetic flux surface at the
radius of interest is given by 
\begin{equation} 
v_{Br0} = c E_{\phi 0}/B_z  
\label{eq:v_Br} 
\end{equation}
\citep[e.g.][]{NNU02}. The magnetic flux speed $v_{Br0}$ cannot be fixed
arbitrarily \citep[e.g.][]{OL01} and is, in general, an eigenvalue of
the global disc/wind problem. However, in the radially localized
formulation of this paper we write $v_{{\rm B}r0} \equiv - \epsilon_{\rm B}
c_{\rm s}$ and treat $\epsilon_{\rm B}$ as a free parameter (see
Section~\ref{subsec:param}).

\subsection{Radial component of the induction equation}
\label{subsec:induc_r}

\begin{eqnarray}
\delt{B_r} & =&  \frac{\partial}{\partial z}(c E_\phi)\label{eq:ind_r} \\
&=& \vr\delz{B_z} + B_z\delz{\vr} - \vz\delz{B_r} - B_r\delz{\vz} +
c \frac{\partial E_{\phi}^\prime}{\partial z}\, .\nonumber
\end{eqnarray}
The radial and vertical components of the induction equation can be
combined into a single relation by using the poloidal magnetic flux
function $\Psi$ defined in Section~\ref{subsec:induc_z}. In view of the
solenoidal condition on $\B$ (equation~\ref{eq:divB}), $\Psi$ satisfies
\begin{displaymath}
B_z=\frac{1}{r}\frac{\partial \Psi}{\partial r}\ , \qquad B_r
= -\frac{1}{r}\frac{\partial \Psi}{\partial z}\ .
\label{eq:Psi}
\end{displaymath}
Hence one can replace equations~(\ref{eq:ind_z}) and~(\ref{eq:ind_r}) by
\begin{equation}
\frac{\partial \Psi}{\partial t} = - r c E_\phi \, ,
\label{eq:Psi_t}
\end{equation}
which implies that, if $\partial B_z/\partial t \ne 0$ (so that $E_\phi
\ne 0$) then, strictly, $\partial B_r/\partial t$ need not vanish. We
will, however, assume (as was done in WK93) that $\partial B_r/\partial
t$ is identically zero. This approximation may be justified on the
grounds that a nonzero value of $B_r$ can be established between the
midplane (where $B_r=0$) and the top of the disc on the ``short''
dynamical time $\tau_{\rm d} \equiv r/v_\phi \approx h/c_{\rm s}$ (where
$h$ is again the disc scale-height) by, for example,
the vertical shear in the radial velocity field acting on $B_z$ (consider,
respectively, the terms $\vz\partial{B_r}/\partial z$ and
$B_z\partial{\vr}/\partial z$ in equation~\ref{eq:ind_r}). Although a
fully self-consistent solution in which $B_r$ at the disc surface
exactly matches the value determined by the global magnetic field
distribution outside the disc (see Section~\ref{subsec:param}) may still
require a contribution from the slow radial diffusion of the poloidal
field, a significant change in the value of $B_r$ at any given radial
location can potentially occur on a time much shorter than the accretion
time. This situation contrasts with that for the vertical field
component $B_z$ (see equations~\ref{eq:ind_z} and~\ref{eq:E_phi}) and is
essentially a consequence of the assumed geometrical thinness of the
disc. An effectively equivalent approximation was adopted in the
\citet{OL01} disc model.

Setting $\partial B_r/\partial t = 0$ in equation~(\ref{eq:ind_r})
implies that $E_\phi$ is constant with height and hence
(using equations~\ref{eq:E_phi} and~\ref{eq:v_Br}) that
\begin{equation}
\label{eq:E_phi/v_B}
cE_{\phi}^{\prime} = v_z B_r + (v_{Br0}-v_r) B_z 
\end{equation}
at any given radius.

\subsection{Azimuthal component of the induction equation}
\label{subsec:induc_phi}

\begin{eqnarray}
\delt{B_{\phi}} &=& \frac{\partial}{\partial z}(\vf B_z - 
\vz B_{\phi}) - \frac{\partial}{\partial r}(\vr B_{\phi} - \vf B_r)
\nonumber\\
&&- c\,\left (\delz{E_r^{\prime}} - \delr{E_z^{\prime}}\right ) \, .
\label{eq:induc_phi}
\end{eqnarray}
In analogy with the case of the radial field component discussed in
Section~\ref{subsec:induc_r}, it can be argued that the
explicit-time-derivative term $\partial{B_{\phi}}/\partial t$ in
equation~(\ref{eq:induc_phi}), which describes variations in $B_\phi$
over the ``long'' radial accretion time, can be neglected in comparison
with the vertical advection term $\vz\partial B_\phi/\partial z$,
which indicates that a measurable azimuthal field component can be
established over the ``short'' dynamical time, in this case primarily
through the radial shear in the azimuthal velocity field acting on $B_r$
(the term $B_r\partial \vf /\partial r$ in
equation~\ref{eq:induc_phi}). We thus set $\partial B_\phi/\partial t =
0$. In addition, we adopt the following simplifications for the
radial-derivative terms:

\begin{displaymath}
\delr{E_z^{'}} \approx 0\, , \qquad
\delr{(\vr B_{\phi})} \approx 0 \, ,
\end{displaymath}

\begin{displaymath}
\delr{\vf} \approx - \frac{\vk}{2r}
\Longrightarrow
\delr{(\vf B_r)} \approx - \frac{3}{2} \frac{B_r \vk}{r} - \vf \delz{B_z}
\end{displaymath}
(see Section~\ref{subsec:model}), where we used $\div{\B} = 0$ in the
second expression. However, for consistency with the rest of our
derivation, we continue to neglect the vertical variation in $B_z$. With
these approximations, and relating $E_r^\prime$ to $E_r$ through
equation~(\ref{eq:E}), we obtain
\begin{equation}
\frac{d}{dz} (c E_r) = -\frac{3}{2} \frac{B_r \vk}{r} \, .
\end{equation}

\subsection{Amp\`ere's Law}
\label{subsec:amp}

The radial and azimuthal components of Amp\`ere's law are, respectively,
\begin{equation}
J_r = - \frac{c}{4 \pi} \frac{d B_{\phi}}{dz} 
\end{equation}
and
\begin{equation}
J_{\phi} = \frac{c}{4\pi} \frac{d B_r}{dz} \, ,
	\label{eq:phi_ampere}
\end{equation}
where we neglected the radial derivative term  ($\partial B_z/\partial
r$) in the last expression. The vertical component is
\begin{equation}
J_z = \frac{c}{4 \pi r}\left[ B_{\phi} + r \delr{B_{\phi}}\right]
\approx 0 \, .
	\label{eq:j_z}
\end{equation}
Our neglect of $J_z$ inside the disc is motivated by the fact that,
under the thin-disc approximation, the magnitudes of the $\J\cross\B$
terms that include this component will be small in comparison with the
terms that involve $J_r$ and $J_\phi$.

\subsection{Ohm's Law}
\label{subsec:ohm}

To simplify this equation, we begin by expressing the 
components of $\E'$ parallel and perpendicular to $\B$ as
\begin{equation}
\E^\prime_{\parallel} = (\E^\prime \bdot \Bh) \Bh =  y \B 
	\label{eq:epar}
\end{equation}
and
\begin{equation}
\E^\prime_{\perp} = - \frac{1}{B^2} (\E^\prime \cross \B) \cross \B =  
- (y \B - \E^\prime) \, ,
	\label{eq:eperp}
\end{equation}
respectively, where $y \equiv \E^\prime \bdot \B/B^2$, $B^2 = \Br^2 + \Ba^2 +
\Bz^2$ and $\Bh$ is the unit vector in the direction of
$\B$. Substituting equations~(\ref{eq:epar}) and~(\ref{eq:eperp}) into
equation~(\ref{eq:J-E}) and using $J_z \approx 0$, we obtain 
\begin{equation}
J_r = y (\sigma_{\rm O} - \sigma_{\rm P}) B_r + 
\frac{\sigma_{\rm H}}{B} (\Ez \Ba - \Ef \Bz) + \sigma_{\rm P} \Er \, ,
	\label{eq:jr1} 
\end{equation}
\begin{equation}
J_{\phi} = {y} (\sigma_{\rm O} - \sigma_{\rm P}) \Ba + 
\frac{\sigma_{\rm H}}{B} (\Er \Bz - \Ez \Br) + \sigma_{\rm P} \Ef \, ,
	\label{eq:jf1}
\end{equation}
\begin{eqnarray}
\lefteqn{\Ez = \frac{-B_z(\Er B_r + \Ef B_{\phi})(\sigma_{\rm O} - 
\sigma_{\rm P})}{B_z^{2} (\sigma_{\rm O} - \sigma_{\rm P})  + B^2
\sigma_{\rm P}} + {} } \nonumber\\ 
& & {} \frac{\sigma_{\rm H} B (\Er B_{\phi} - \Ef B_r)}{B_z^{2} 
(\sigma_{\rm O} - \sigma_{\rm P})  + B^2 \sigma_{\rm P}} \ .
	\label{eq:ez1}
\end{eqnarray}

\subsection{System of ODEs in $z$}
	\label{subsec:system}

Summarizing, after the simplifications applied in
Section~\ref{subsec:ODE}, we have the following system of nonlinear,
ordinary differential equations in $z$:
\begin{equation}
\rho v_z \frac{d v_r}{dz} +  \frac{2 \rho \vk}{r} (\vk - v_{\phi}) = 
\frac{1}{c}( J_{\phi} B_z) \, ,
	\label{eq:r_motion}
\end{equation}
\begin{equation}
\rho v_z \frac{d v_{\phi}}{dz} + \frac{\rho v_r \vk}{2r} = 
-\frac{1}{c}(J_r B_z) \, ,
	\label{eq:phi_motion}
\end{equation}
\begin{equation}
(c_{\rm s}^2 - v_z^2)\frac{d\rho}{dz} + \frac{\rho \vk^2}{r}\frac{z}{r} = 
\frac{1}{c}(J_r B_{\phi} - J_{\phi} B_r) \, ,
	\label{eq:z-motion}
\end{equation}
\begin{equation}
\frac{d}{dz}(cE_r) = -\frac{3}{2} \frac{B_r \vk}{r} \, , 
	\label{eq:phi_induction}
\end{equation}
\begin{equation}
J_r = - \frac{c}{4 \pi} \frac{d B_{\phi}}{dz} \, ,
	\label{r_ampere}
\end{equation}
\begin{equation}
J_{\phi} = \frac{c}{4\pi} \frac{d B_r}{dz} \, .
	\label{eq:phi_ampere1}
\end{equation}

\subsection{Normalized Equations}
	\label{subsec:dimensionless}

Equations (\ref{eq:r_motion})--(\ref{eq:phi_ampere1}) form a system of
ordinary differential equations in $\vr$, $\vf$, $\rho$, $E_r$, $\Ba$
and $\Br$. The remaining variables -- $\vz$, $J_r$, $J_{\phi}$, $\Er$,
$\Ef$ and $\Ez$ -- may be found algebraically via equations (\ref{eq:E})
and (\ref{eq:jr1})--(\ref{eq:ez1}), together with the condition $\rho
v_z = \rm{const}$. Furthermore, as discussed above, under the adopted
approximations $B_z$ and $E_{\phi}$ are constant with height and $J_z
\approx 0$.

These equations can be expressed in dimensionless form by normalizing
the variables as follows:
\begin{equation}
\label{eq:nondim1}
\zt \equiv \frac{z}{h_{\rm T}}\ , \qquad
\rhot \equiv \frac{\rho(r,z)}{\rho_0(r)}\ , 
\end{equation}
\begin{equation}
\label{eq:nondim2}
\w \equiv \frac{\v - \vk \fh}{c_{\rm s}}\ , \
\we \equiv  \frac{c \E/B_0 + \vk \rh}{c_{\rm s}}\ , \
\e^\prime = \frac{c \E^\prime}{c_{\rm s} B_0} \ ,
\end{equation}
\begin{equation}
\label{eq:nondim3}
\j = \frac{4 \pi h_{\rm T} \J}{c B_0}\ , \quad \bmath{\tilde \sigma} =
\frac{4 \pi h_{\rm T} c_{\rm s}\bmath{\sigma}}{c^2}\ , \quad \b =
\frac{\B}{B_0} \ ,
\end{equation}
where $h_{\rm T} = c_{\rm s} r/ \vk$ is the tidal scale-height of the
disc. Note that, under the adopted field symmetry, $B_0 = B_z$ and
hence $\bz = 1$. The quantity $\w$ represents the
normalized (by $\cs$) fluid velocity in a frame that rotates with the
local Keplerian angular velocity $\Omega_{\rm K} = \vk/r$. The quantity
$\we$ is an analogously reduced effective flux-surface velocity. In
particular, note that $w_{{\rm E}\phi 0} = v_{Br0}/\cs$ (see
equation~\ref{eq:v_Br}). One can similarly express the effective
midplane angular velocity of the magnetic flux surfaces as $\Omega_{{\rm
B}0} = -cE_{r0}/rB_0$, which implies that $w_{{\rm E}r0} = - r
(\Omega_{{\rm B}0}-\Omega_{\rm K})/\cs$.

With these normalizations, the following dimensionless system of
equations is obtained:
\begin{equation}
\frac{d w_r}{d\zt} = \frac{1}{\wz}\left[\frac{a_0^2}{\rhot} \jf + 2
\wf\right]\ ,
	\label{eq:r_motiond}
\end{equation}
\begin{equation}
\frac{d w_{\phi}}{d\zt} = -\frac{1}{w_z} \left[\frac{a_0^2}{\rhot} \jr +
\frac{w_r}{2}\right]\ ,
	\label{eq:phi_motiond}
\end{equation}
\begin{equation}
\frac{d \ln{\rhot}}{d\zt} = \frac{1}{1 - w_z^2} \left[\frac{a_0^2}{\rhot} (\jr
\ba - \jf \br) - \zt \right]\ ,
	\label{eq:z_motiond}
\end{equation}
\begin{equation}
\frac{d w_{{\rm E}r}}{d\zt} = -\frac{3}{2} \br \, , 
	\label{eq:phi_inductiond}
\end{equation}
\begin{equation}
\frac{d \br}{d\zt} = \jf \, ,
	\label{eq:phi_ampered}
\end{equation}
\begin{equation}
\frac{d \ba}{d\zt} = - \jr \, ,
	\label{eq:r_ampered}
\end{equation}
\begin{equation}
\er = w_{{\rm E}r} + \wf - \wz\ba \,  ,
	\label{eq:E-E'}
\end{equation}
\begin{equation}
\ef = -\epsilon_{\rm B} + w_z \br - w_r \, , 
	\label{eq:E-E'1}
\end{equation}
\begin{equation}
\jr = y (\sigpar - \sigP) \br + 
\frac{\sigH}{b} (\ez \ba - \ef) + \sigP \er \, ,
	\label{eq:jr}
\end{equation}
\begin{equation}
\jf = y (\sigpar - \sigP) \ba + 
\frac{\sigH}{b} (\er - \ez \br) + \sigP \ef \, ,
	\label{eq:jphi} 
\end{equation}
\begin{eqnarray}
\lefteqn{\ez = \frac{-(\er \br + \ef \ba)(\sigpar - 
\sigP)}{ (\sigpar - \sigP)  + b^2 \sigP} + {} }
\nonumber\\ 
& & {} \frac{\sigH b (\er \ba - \ef \br)}{
(\sigpar - \sigP)  + b^2 \sigP} \ .
	\label{eq1:ez}
\end{eqnarray}

\noindent
In the above expressions $a_0$ is the midplane value of $a \equiv
v_{{\rm A}}/c_{\rm s}$, where the Alfv\'en speed is given by
\begin{equation}
v_{\rm A} = \frac{|\mathbf{B}|}{\sqrt {4 \pi \rho}}\ ,
	\label{eq:alfven}
\end{equation}

\noindent
and $\epsilon_{\rm B}
\equiv -w_{{\rm E}\phi}= -v_{Br0}/\cs$ is the normalized value of the
vertically uniform $E_\phi$.

Note in this connection that $w_{{\rm E}r}$, the radial component of the
reduced electric field, is expected to be $> 0$ below the base of the
wind. This was demonstrated in WK93 for the ambipolar diffusion case
and is also true in the Hall and Ohm regimes. The reason for this is
that the matter inside the disc rotates at sub-Keplerian speeds because
it loses angular momentum to the field, and since the motion of the
field lines is controlled by that of the matter (given that the
quasi-hydrostatic layer, where the bulk of the mass is located, is
thermal pressure-dominated), $\Omega_{\rm B}$ must remain $\lesssim
v_\phi/r$. Hence $w_{{\rm E}r} = - r (\Omega_{\rm B}-\Omega_{\rm
K})/\cs$ would be $>0$. As the magnetic field bends away from the disc
rotation axis ($B_r > 0$ above the midplane), $\Omega_{\rm K}$ decreases
with distance along the field line, resulting in a corresponding decline
in $w_{{\rm E}r}$. (Equivalently, equation~\ref{eq:phi_inductiond} shows
that $w_{{\rm E}r}$ decreases with $\zt$ for $\br>0$.) Eventually
$w_{{\rm E}r}$ vanishes at some height above the midplane, which roughly
coincides with the location of the base of the wind (see
Section~\ref{subsec:base}).

As the height of the sonic point (denoted by a subscript s) and the gas
density at that location (or, equivalently, the vertical velocity at the
midplane, given that $\rho v_z$ is constant with height) are not known a
priori, we treat $\zt_{\rm s}$ and $w_{{z}\rm 0}$ as additional
variables, which satisfy
\begin{equation} 
\frac{d \zt_{\rm s}}{d\zt} = 0
\label{eq:z_s} 
\end{equation}
and
\begin{equation} 
\frac{d w_{z 0}}{d\zt} = 0 \, .
\label{eq:w0} 
\end{equation}
This allows us to find the position of the sonic point and the upward
mass flux self-consistently (see Paper~II).

\subsection{Parameters}
	\label{subsec:param}

The following parameters characterize the disc solutions in our model:
\begin{enumerate}

\item $a_0 \equiv v_{{\rm A}0}/c_{\rm s}$, the midplane ratio of
the Alfv\'en speed (based on the large-scale vertical field component) to
the isothermal sound speed. This parameter measures the strength of the
ordered magnetic field that threads the disc.

\item $c_{\rm s}/\vk = h_{\rm T}/r$, the ratio of the tidal
scale-height to the disc radius. While this parameter, which measures the
geometric thinness of the disc, does not appear explicitly in the
normalized equations, it nevertheless serves to constrain physically
viable solutions (Section~\ref{sec:constraints}) and is used in matching
the disc solutions to the self-similar wind solutions (see Paper~II).

\item The midplane ratios of the conductivity tensor components:
$[\sigma_{\rm P}/\sigma_\perp]_0$ (or $[\sigma_{\rm H}/\sigma_\perp]_0$)
and $[\sigma_\perp/\sigma_{\rm O}]_0$. They characterize the
conductivity regime of the fluid (Section \ref{subsec:model}). In
general the conductivity tensor components vary with height, reflecting
the ionization structure of the disc \citep[e.g.][]{SW05}. However, the
explicit solutions derived in Paper~II correspond to a simplified
ionization prescription wherein the above ratios remain constant with
$z$.

\item The midplane Elsasser number $\Lambda_0 \equiv v_{{\rm
A}0}^{2}/(\Omega_{\rm K}\eta_{\perp 0})$, where $\eta_\perp\equiv
c^2/4\pi \sigma_\perp$ is the ``perpendicular'' magnetic
diffusivity. This parameter measures the degree of coupling between the
neutrals and the magnetic field, with values $\gg 1$ and $\ll 1$
corresponding to strong and weak coupling, respectively.\footnote{The
Elsasser number $\Lambda=v_{\rm A}^{2}/(\Omega_{\rm K}\eta_{\perp})$ is
distinct from the Lundquist number $S\equiv v_{\rm A} L/\eta_{\rm O})$
and from the magnetic Reynolds number Re$_{\rm M}\equiv
V L/\eta_{\rm O}$ (where $V$ and $L$ are characteristic speed and
length-scale, respectively), which have been used in similar contexts in
the literature. This quantity was labeled by the symbol $\chi$ in the
MRI linear stability analyses of \citet{War99} and \citet{SW03,SW05},
where, in fact, its general form (which allows for variation with
height) was considered.}

\item $\epsilon \equiv -v_{r0}/c_{\rm s}$, the normalized inward radial
speed at the midplane. This is a free parameter of the \emph{disc
solution}; its value is determined (for given values of the other
parameters) at the step where we match it to the BP82 self-similar
global wind solution by imposing the Alfv\'en critical-point
constraint on the wind (see Paper~II). Although the
value of $\epsilon$ could in principle be negative (as it in fact is in
certain viscous disc models; e.g. \citealt{TL02}), in the context
of our model formulation we generally expect $\epsilon > 0$ for
physically viable, exclusively wind-driving discs.

\item $\epsilon_{\rm B} \equiv -v_{Br0}/c_{\rm s} = -cE_{\phi 0}/c_{\rm
s} B_z$, the normalized azimuthal component of the electric field $\E$
(vertically constant by equation~\ref{eq:ind_r}), measures the
radial drift speed of the poloidal magnetic field lines. This
parameter vanishes in a strictly steady-state solution but is nonzero
if the magnetic field lines drift radially on the ``long'' accretion
time-scale $\tau_{\rm a}$.
\end{enumerate} 

In a global treatment of the disc/wind problem one can relate
$\epsilon_{\rm B}$ to the value of $B_r$ at the disc surface. The latter
is determined by the magnetic field distribution outside the disc, and,
in the limit of a potential ($\grad \cross \B = 0$) external field, can
be inferred from the radial distribution of $B_z$ along the disc surface
\citep[e.g.][]{CM93,LPP94,KK02}. In the numerical solutions presented in
Paper~II we simplify the calculation by setting $\epsilon_{\rm B}=0$;
this condition is then used to determine the value of $B_r$ at the disc
surface. WK93, who employed a similar radially localized model, derived
solutions for both positive and negative values of $\epsilon_{\rm B}$
(subject to the physically motivated constraint $\epsilon_{\rm B} <
\epsilon$) and showed that disc configurations with the same value of
$(\epsilon - \epsilon_{\rm B})$ were very similar to each other. This
result suggests that setting $\epsilon_{\rm B}$ equal to zero should not
strongly impact the generality of the results. Further discussion of
this approximation is given in Appendix~\ref{sec:appA}.

\subsection{Comparison with the multi-fluid approach}
\label{subsec:multi}

A commonly used alternative to the conductivity-tensor formalism is
based on writing down the separate equations of motion for each
charged-particle species (e.g. WK93). In this case, instead of employing
two independent ratios of the conductivity-tensor components, the fluid
can be characterized by the electron and ion Hall parameters. Indeed, we
already noted in Section~\ref{sec:formulation} how the distinct
conductivity regimes of the fluid can be delineated directly in terms of
$\beta_{\rm e}$ and $\beta_{\rm i}$. It is also instructive to evaluate
the limiting forms of the magnetic coupling parameter $\Lambda$ -- the
``third parameter'' in our classification scheme of viable solutions
(see Section~\ref{sec:constraints}) -- in the different conductivity
regimes. In the ambipolar diffusion limit $|\beta_{\rm e}| \gg
|\beta_{\rm i}| \gg 1$ and hence
\begin{equation}
\sigma_{\perp} \approx \sigma_{\rm P} \approx \frac{c e}{B} \frac{n_{\rm
e}}{\beta_{\rm i}}
\end{equation}
(see equations~\ref{eq:sig3} and~\ref{eq1:sigperp}). In this case
$\Lambda$ reduces to
\begin{equation}
\Upsilon \equiv \frac{\gamma_{\rm i} \rho_{\rm i}}{\Omega_{\rm K}}\ ,
\label{eq:eta}
\end{equation}
the ratio of the Keplerian rotation time to the neutral--ion
momentum exchange time. This parameter has emerged as the natural
measure of the field--matter coupling in the ambipolar
diffusion-dominated disc model of WK93 (where it was labelled $\eta$),
as well as in studies of the linear \citep[e.g.][]{BB94} and the nonlinear
\citep[e.g.][]{Mac95,Bra95,HS98} evolution of the MRI in such
discs. Indeed, since the ions are well coupled to the field in this
limit ($|\beta_{\rm i}| > 1$), the neutrals will be well coupled to the
field if their momentum exchange with the charged particles (which is
dominated by their interaction with the comparatively massive ions)
occurs on a time-scale that is short in comparison with the dynamical
time (corresponding to $\Upsilon > 1$).

On the other hand, in the Hall regime $|\beta_{\rm e}| \gg 1$ and
$|\beta_{\rm i}| \ll 1$, which implies
\begin{equation}
\sigma_{\perp} \approx |\sigma_{\rm H}| \approx  \frac{c e n_{\rm
e}}{|B|}
\end{equation}
(see equations~\ref{eq:sig2} and~\ref{eq1:sigperp}) and hence $\Lambda =
\Upsilon|\beta_{\rm i}|$. The parameter combination $\Upsilon
|\beta_{\rm i}|$, which also figures prominently in linear
\citep[e.g.][]{War99,BT01} and nonlinear \citep[e.g.][]{SS02a,SS02b}
studies of disc MRI in the Hall limit, similarly has a clear physical
meaning. Indeed, in contradistinction to the ambipolar diffusion regime,
the ions are {\em not} well coupled to the field in this case
($|\beta_{\rm i}|< 1$). In order for the neutrals to be well coupled to
the field in the Hall regime it is, therefore, not sufficient for them
to be well coupled to the ions ($\Upsilon > 1$); rather, the product
$\Upsilon |\beta_{\rm i}|$ must be $>1$ in this case.

Finally, in the Ohm regime, where even the ``electrons'' are not
well-coupled to the magnetic field ($|\beta_{\rm e}| \ll 1$),
\begin{equation} \sigma_{\perp} \approx \sigma_{\rm P} \approx \frac{c e
n_{\rm e}}{B} \beta_{\rm e} \end{equation} and therefore $\Lambda =
\Upsilon \beta_{\rm e} \beta_{\rm i}$, implying a further tightening of
the requirement for good coupling between the neutrals and the field.

As we show in Sections~\ref{sec:constraints}
and~\ref{sec:Ohm_constraints}, the parameter $\Upsilon$ plays a
fundamental role in the theory of wind-driving discs. In fact, it turns
out that the condition $\Upsilon \gtrsim 1$ (which, according to the
argument given above, signifies good coupling of the neutrals to the
field only in the ambipolar diffusion regime) must be satisfied by
\emph{all} viable configurations of this type, irrespective of the conductivity
domain that characterizes the disc. We elaborate on this point in
Section~\ref{sec:discuss}.

One can use the definitions of the conductivity components to express
their ratios in terms of $\beta_{\rm e}$ and $\beta_{\rm
i}$. Specializing to the case $q \ll 1$, we have
\begin{equation}
\frac{\sigma_{\rm P}}{\sigma_{\rm H}} = \frac{1 + \beta_{\rm e}
\beta_{\rm i}}{\beta_{\rm e}}\ .
\label{eq:sHP}
\end{equation}
Furthermore, in the Hall and Ohm limits we get
\begin{equation}
\frac{\sigma_{\rm O}}{\sigma_{\rm H}} =  \left\{
\begin{array}{lcc}
\beta_{\rm e} \quad \ \ \ \ {\rm Hall,} \\
\beta_{\rm e}^{-1} \quad \ \ {\rm Ohm.}
\end{array} \right.
\label{eq:spar}
\end{equation}
We also have
\begin{equation}
\tilde{\sigma}_{\perp 0}
= \frac{\Lambda_0}{a_0^2} \ ,
\label{eq:bi}
\end{equation}
which follows from the definition of the Elsasser number.

\section{Parameter Constraints}
\label{sec:constraints}

As previously shown by WK93 and \citet{Kon97}, viable wind-driving disc
solutions in the strong-coupling regime (defined by $\Lambda_0$ not
being $\ll 1$) occupy a limited region of parameter space. This region
is determined by the requirements that (i) the flow remain sub-Keplerian
within the disc, (ii) a wind is driven from the disc surface (i.e. a
wind launching criterion is satisfied), (iii) only the upper layers of
the disc participate in the outflow and (iv) the rate of heating by
Joule dissipation is bounded by the rate of gravitational energy
release.  In this section we generalize these conditions, which were
originally derived for discs in the ambipolar diffusion regime, and
apply them to discs in the Hall conductivity domain. The corresponding
constraints for discs in the Ohm regime are presented in
Section~\ref{sec:Ohm_constraints}. The viable solution regions can in
general be delineated by the midplane values of two independent
conductivity-tensor components and of the Elsasser number $\Lambda$ (see
items ($iii$) and ($iv$) in Section~\ref{subsec:param}). While we employ
this representation in Paper~II, in this paper we use instead the
parameters $\psi \equiv \Upsilon_0$, $\beta \equiv 1/\beta_{\rm i 0}$
and $q \beta = 1/ \beta_{\rm e 0}$, which are appropriate to the
two-component plasma to which we specialize and afford
additional insights into the problem. This choice also allows us to make
direct comparisons with the results obtained in WK93 for the ambipolar
diffusion regime.

The relevant constraints are derived by focussing, as
in WK93, on the quasi-hydrostatic disc region adjacent to the midplane,
where $\wz^2 \ll 1$. In this limit the vertical momentum conservation
equation~(\ref{eq:z_motiond}) reduces to
\begin{equation}
\frac{d \rhot}{d\zt} \approx a_0^2 (\jr \ba - \jf \br) - \rhot \zt \, .
\label{eq:z_motiond1}
\end{equation}
The last term on the right-hand side of this equation, which
represents tidal compression, can generally be neglected in
comparison with the magnetic squeezing term (see
Section~\ref{sec:discuss}). Applying this approximation and substituting
for $\jf$ and $\jr$ from equations~(\ref{eq:phi_ampered})
and~(\ref{eq:r_ampered}), respectively, one can then integrate
equation~(\ref{eq:z_motiond1}) to deduce that
\begin{equation}
\rhot + \frac{a_0^2}{2} (b_r^2 + b_{\phi}^2) \approx {\rm const}
\end{equation}
within the disc and, therefore, that
\begin{equation}
b_{r{\rm b}}^2 + b_{\phi{\rm b}}^2 \approx \frac{2}{a_0^2}
\label{eq:rphi_B}
\end{equation}
at the base of the wind (subscript b; the effective surface of the
disc), where $\rhot \ll 1$. Within the quasi-hydrostatic layer it is
appropriate to set $\rhot \approx 1$. The two other components of the
momentum conservation relation (equations~\ref{eq:r_motiond}
and~\ref{eq:phi_motiond}) then reduce to
\begin{equation}
\wf \approx - \frac{a_0^2}{2} \jf
\label{eq:r_alg}
\end{equation}
and
\begin{equation}
\wr \approx -2 a_0^2 \jr\, .
\label{eq:phi_alg}
\end{equation}

In the remainder of this section (as well as in
Section~\ref{sec:Ohm_constraints})
we assume that the ion mass density $\rho_{\rm i}$ and the collisional
coupling coefficient $\gamma_{\rm i}$ (equation~\ref{eq:gamma}) are
constant with height in the quasi-hydrostatic zone. Under this
assumption we can neglect the $z$-variation of the quantity $\Upsilon$
within this layer and set it equal to $\Upsilon_0=\psi$.\footnote{In WK93
it was assumed that $\rho_{\rm i}$ and $\gamma_{\rm i}$ are constant
from $z=0$ all the way to the sonic point of the wind, and it was noted
that this might be a reasonable approximation to the actual conditions
in some protostellar discs on scales $\gtrsim 10^2\,$AU, where ambipolar
diffusion dominates. In the more general case considered in this paper,
where we incorporate also the Hall and Ohm diffusivity regimes that are
relevant at higher densities, a similar assumption would not be
realistic. However, here we only make this approximation in the
quasi-hydrostatic layer, where it helps to simplify the analysis.}

\subsection{Sub-Keplerian flow below the launching region}
\label{subsec:subK}

We seek to identify the regions of parameter space where $\wf< 0$ below
the launching region, as expected for physically viable solutions (see
Section~\ref{subsec:dimensionless} and WK93). We begin by expressing
$\jr$ and $\jf$ as functions of $\br$, $\ba$, $w_{{\rm E}r}$ and our
model parameters by substituting equations~(\ref{eq:E}),
(\ref{eq:r_alg}) and (\ref{eq:phi_alg}) into
equation~(\ref{eq1:E-J}). Equation~(\ref{eq:r_alg}) can then be used to
deduce the condition that must be satisfied for the flow to remain
sub-Keplerian within the disc. The radial and azimuthal components of
equation (\ref{eq1:E-J}) are, respectively,
\begin{eqnarray}
\lefteqn{\left[ b^2 \frac{\sigma_\perp}{\sigma_{\rm O}} +
\left(\frac{\sigma_{\rm P}}{\sigma_\perp} -
\frac{\sigma_\perp}{\sigma_{\rm O}}   \right)  (1 + \ba^2) \right] \jr +
{} }   \nonumber \\
\vspace{10pt}
& & {} \left[ b^2 \frac{\Lambda_0}{2} + b\frac{\sigma_{\rm
H}}{\sigma_\perp} - \left(\frac{\sigma_{\rm P}}{\sigma_\perp} -
\frac{\sigma_\perp}{\sigma_{\rm O}}   \right) \br \ba \right] \jf =
b^2\frac{\Lambda_0}{a_0^2} w_{{\rm E}r} 
\label{eq:Er}
\end{eqnarray}
and
\begin{eqnarray}
\lefteqn{\left[ 2 b^2 \Lambda_0 + b \frac{\sigma_{\rm H}}{\sigma_\perp} +
\left(\frac{\sigma_{\rm P}}{\sigma_\perp} -
\frac{\sigma_\perp}{\sigma_{\rm O}}   \right) \br \ba \right] \jr}  
\nonumber \\
\vspace{10pt}
& & \mbox{ } 
- \left[ b^2 \frac{\sigma_\perp}{\sigma_{\rm O}} +
\left(\frac{\sigma_{\rm P}}{\sigma_\perp} -
\frac{\sigma_\perp}{\sigma_{\rm O}}   \right) (1 + \br^2) \right] \jf =
b^2 \frac{\Lambda_0}{a_0^2} \epsilon_{\rm B} \, . 
\label{eq:epsilonB}
\end{eqnarray}
These relations can be cast in terms of $\br$, $\ba$, $w_{{\rm E}r}$ and
the parameters $\psi$, $\beta$ and $q$, which are the variables employed
by WK93 in their analysis of the ambipolar diffusion
regime. Specifically, inverting equations (\ref{eq:Er}) and
(\ref{eq:epsilonB}) and substituting
\begin{equation}
b\frac{\sigma_{\rm H}}{\sigma_\perp} = b^2 \left(\frac{1 - q}{1 + q}
\right) \frac{\beta \Lambda_0}{\psi} \ ,
\label{eq:sigmaratio1}
\end{equation}
\begin{equation}
b^2 \frac{\sigma_\perp}{\sigma_{\rm O}} = b^2 \frac{q
\beta^2}{1 + q}\frac{\Lambda_0}{\psi}
\label{eq:sigmaratio2}
\end{equation}
and
\begin{equation}
\left(\frac{\sigma_{\rm P}}{\sigma_\perp} -
\frac{\sigma_\perp}{\sigma_{\rm O}}   \right) =
\frac{b^2}{1+q}\frac{\Lambda_0}{\psi} \ ,
\label{eq:sigmaratio3}
\end{equation}
we find
\begin{equation}
\left( \begin{array}{c}
\jr \\
\\
\jf 
\end{array} \right)
 = \frac{\psi}{a_0^2} \frac{(1+q)}{K}\left( \begin{array}{cc}
A_1 & A_2 \\
\\
A_3 & A_4 
\end{array} \right) \left( \begin{array}{c}
w_{{\rm E}r}\\
\\
\epsilon_{\rm B}
\end{array} \right) \ .
	\label{eq1:matrix}
\end{equation}

\vspace{10pt}

\noindent
In these expressions
\begin{equation}
A_1 = 1 + q\beta^2 + \br^2 \, ,
\label{eq:A1}
\end{equation}
\begin{equation}
A_2 = \frac{\psi}{2}(1+q) + \beta(1-q) - \br \ba \, ,
\label{eq:A2}
\end{equation}
\begin{equation}
A_3 = 2 \psi (1+q) + \beta(1-q) + \br \ba \, ,
\label{eq:A3}
\end{equation}
\begin{equation}
A_4 = -(1 + q\beta^2 + \ba^2)
\label{eq:A4}
\end{equation}
and
\begin{eqnarray}
\lefteqn{K = (1 + q\beta^2)(\br^2 + \ba^2) + (1 + \beta^2)(1 + q^2
\beta^2)} \nonumber \\ 
& & \mbox{ } + \psi(1+q)\left[  \frac{5}{2}\beta (1-q) 
- \frac{3}{2} \br \ba + \psi (1+q)\right] \, .
\label{eq:K}
\end{eqnarray}

In the remainder of this paper we assume, for simplicity, that
$\epsilon_{\rm B} \approx 0$ (see Section~\ref{subsec:param} and
Appendix~\ref{sec:appA}). Substituting $\jr$ from
equation~(\ref{eq:phi_alg}) into equation~(\ref{eq1:matrix}) and using
the fact that $w_{{\rm E}r} > 0$ (see Section
\ref{subsec:dimensionless}) and $\wr < 0$ below the disc surface, we
infer that $K$ must be $>0$ within the disc. We can also substitute
$\jf$ from equation~(\ref{eq1:matrix}) into equation~(\ref{eq:r_alg}) to
obtain an explicit expression for the sub-Keplerian flow condition:
\begin{equation}
\wf \approx -\frac{\psi (1 + q)}{2} \frac{[2 \psi (1 + q) + \beta (1 -
q)  + \br \ba] w_{{\rm E}r} }{K}  < 0 \, .
\label{eq:subK}
\end{equation}
In the rest of this section and in Section~\ref{sec:Ohm_constraints}
we also specialize to the case
$q \ll 1$ (or, equivalently, $1\pm q \approx 1$), which provides a
good approximation for discs in which grains are not the dominant
charge carriers. It is straightforward to carry out this analysis in the
other case of special interest, namely $q=1$, which arises when grains
dominate both the ``+'' and the ``-'' charged components (see
Section~\ref{sec:discuss}). 
In the limit $q \ll 1$, the expression for $\wf$ reduces to
\begin{eqnarray}
\label{eq:wf}
\lefteqn{\wf\approx} \\
\lefteqn{\frac{-\frac{1}{2}\psi (2 \psi + \beta + \br \ba) w_{{\rm E}r} }{(1 +
q\beta^2)(\br^2 + \ba^2) + (1 + \beta^2)(1 + q^2 \beta^2)
+ \psi (\frac{5}{2}\beta - \frac{3}{2} \br \ba + \psi) }\, . \nonumber}
\end{eqnarray}

In the ambipolar diffusion limit ($|\beta_{\rm e}| \gg
|\beta_{\rm i}| \gg 1$) we can take $\beta$ ($\equiv 1/\beta_{\rm i0}$),
$q \beta$ ($\equiv 1/\beta_{\rm e 0}$) and $q \beta^2$ ($\equiv
1/\beta_{\rm e 0} \beta_{\rm i 0}$) to be all $\ll 1$. The right-hand
side of equation~(\ref{eq:wf}) then reduces to the right-hand side of
equation~(4.3) in WK93 (where the minus sign in front of the numerator
is a typographical error). On the other hand, in
the Hall limit ($|\beta_{\rm i}| \ll 1 \ll |\beta_{\rm e}|$) one has
$\beta \gg 1$ and $q \beta \ll 1$, but $q \beta^2$ may be either $>1$ or
$<1$. In the Ohm regime ($|\beta_{\rm i}| \ll |\beta_{\rm e}| \ll 1$)
both $\beta$ and $q\beta$ (and hence also $q\beta^2$) are $\gg 1$.

Given that $w_{{\rm E}r} > 0$ and $K>0$ inside the disc, the condition
$\wf < 0$ implies that
\begin{equation}
\frac{d\br}{d\ba} =  - \frac{\jf}{\jr} =  - \frac{2 \psi + \beta +
\br \ba}{1 + q\beta^2 + \br^2}
\label{eq:B_rB_phi}
\end{equation}
(where we used equations~\ref{eq1:matrix}, \ref{eq:A1} and~\ref{eq:A3}
in the limit $q\ll 1$) is $< 0$, or
\begin{equation}
-\br \ba < 2\psi + \beta \equiv C
\label{eq:rphi_B1}
\end{equation}
below the disc surface.
The fact that $\br\ba < 0$ above the midplane in turn implies that $C$
must be $> 0$ and hence that
\begin{equation}
\beta > - 2 \psi\, .
\label{eq:beta}
\end{equation}

The form of equation~(\ref{eq:B_rB_phi}) suggests that the Hall
parameter regime can be subdivided into four sub-zones, depending on how
the ion slip factor $s_0=1/q \beta^2$ and the Elsasser number $\Lambda_0
= \psi/|\beta|$ compare with 1 and $1/2$, respectively \citep{Kon97}. In
Table~\ref{table:constraints}, where we list the parameter constraints
for viable wind-driving disc solutions in the Hall regime, we label
these sub-regions by the numerals $i$ through $iv$.

Combining equations~(\ref{eq:rphi_B}) and~(\ref{eq:rphi_B1}), we infer
that the solutions must satisfy
\begin{equation}
b_{r{\rm b}}^4 - \frac{2}{a_0^2} b_{r{\rm b}}^2 + C^2 > 0
\label{eq:quadBr}
\end{equation}
in order for the accretion flow to remain sub-Keplerian all the way to
the top of the disc. Assuming an equality, we find that solutions only
exist if $a_0 > C^{-1/2}$. This translates into $a_0 > (2
\psi)^{-1/2}$ for Cases~$(i)$ and~$(ii)$ in Table~\ref{table:constraints}
(for which $C \approx 2 \psi$) and $a_0^2 > 1/\beta$ for the
remaining two cases (where $C \approx \beta$). These constraints are
summarized in the first column of inequalities in the
table.\footnote{Note that, although the ion Hall parameter can in principle
assume a negative value, all entries of $\beta_{\rm i0}$ in
Tables~\ref{table:constraints1} and~\ref{table:Ohm_constraints1} that
are not marked as absolute values are in fact $> 0$ (see
Section~\ref{sec:discuss}).} 
 
\setlength{\cellspacetoplimit}{1mm}
\setlength{\cellspacebottomlimit}{1mm}
\begin{table*}
\caption{Parameter constraints for wind-driving disc solutions in the
limit where the Hall diffusivity dominates. Four distinct cases can be
identified, depending on how the values of $s_0 = 1/q\beta^2 = \beta_{\rm e 0}
\beta_{\rm i 0}$ and of $2\Lambda_0 = 2\psi/|\beta| =
2\Upsilon_0 |\beta_{\rm i 0}|$ compare with 1. The first inequality
expresses the requirement that the disc remain
sub-Keplerian below the wind zone ($\tilde{z} < \tilde{z}_{\rm
b}$), the second is the wind launching condition (the requirement that
the magnetic field lines be sufficiently inclined to the vertical for
centrifugal acceleration to occur), the third ensures that the base of
the wind is located above the (magnetically reduced) density
scale-height and the fourth specifies that the rate of Joule heating at
the midplane should not exceed the rate of release of gravitational
potential energy there.}
\label{table:constraints} 
\begin{center}
\begin{tabular}{Sc|Sc|Sc|ScScScScScScScScSc} 
\hline 
{\normalsize Case} & \multicolumn{2}{c}{{\normalsize Limits}} &
\multicolumn{9}{c}{{\normalsize Parameter Constraints -- Hall Limit }}\\ 
& $s_0=\beta_{\rm e 0} \beta_{\rm i 0}$ & $\Lambda_0=\Upsilon_0 |\beta_{\rm i
0}|$ & \multicolumn{9}{c}{{\small (multi-fluid formulation)}}\\ 
\hline 
$i$ & $> 1$ & $> 1/2$ & $(2\Upsilon_0)^{-1/2}$ & $\lesssim$ & $a_0$ &
$\lesssim$ & $2$ & $\lesssim$ & $\epsilon\Upsilon_0$ & $\lesssim$ & $\vk/2
c_{\rm s}$ \\ 
$ii$ & $ > 1$ & $< 1/2$ & $\beti^{1/2}$ & $ \lesssim$ &
$a_0$ & $\lesssim$ & $2 (\Upsilon_0 \beti)^{1/2}$ & $\lesssim$ & $\epsilon/
2\beti$ & $\lesssim$ & $\Upsilon_0 \beti \vk/ c_{\rm s}$ \\ 
$iii$ & $< 1$ & $>
1/2$ & $(2\Upsilon_0)^{-1/2}$ & $\lesssim$ & $a_0$ & $\lesssim$ & $2$ &
$\lesssim$ & $\epsilon \Upsilon_0 \beta_{\rm e0} \beta_{\rm i0}$ & $\lesssim$
& $\vk/2 c_{\rm s}$ \\ 
$iv$ & $< 1$ & $< 1/2$ & $\beti^{1/2}$ & $\lesssim$ & $a_0$ & $\lesssim$
& $2(\Upsilon_0 \beti)^{1/2}$ & $\lesssim$ & $\epsilon \beta_{\rm e0}/2$ &
$\lesssim$ & $\Upsilon_0 \beti \vk/ c_{\rm s}$ \\ 
\hline 
\end{tabular} 
\end{center} 
\end{table*}

\setlength{\cellspacetoplimit}{1mm}
\setlength{\cellspacebottomlimit}{1mm}
\begin{table*}
\caption{Key properties of viable disc solutions in the Hall
regime. Listed, in order, are the midplane values of $|db_r/db_\phi|$,
the magnetically compressed scale-height in units of the tidal
scale-height ($\tilde h \equiv h /h_{\rm T}$), the similarly normalized
vertical location of the base of the wind $\tilde{z}_{\rm b}$ in units
of $\tilde{h}$ and the normalized Joule dissipation rate $\j \bdot
{\bmath e}^\prime$ at the midplane.}
\label{table:constraints1}
\begin{center}
\begin{tabular}{Sc|Sc|Sc|SrScScSc}
\hline
{\normalsize Case} & \multicolumn{2}{c}{{\normalsize Limits}} &
\multicolumn{4}{c}{{\normalsize Solution Characteristics -- Hall Limit }} \\
& $s_0= \bete \beti$ & $\Lambda_0=\Upsilon_0 |\beti|$ & 
$|db_r/db_\phi|_0$ &  $\tilde
h$ & $\tilde z_{\rm b}/\tilde h$ &  
$(\j \bdot {\bmath e}^\prime)_0 $ \\ 
\hline 
$i$ &
$> 1$ & $> 1/2$ &  $2 \Upsilon_0 \ \qquad (> 1)$ & $a_0/ \epsilon
\Upsilon_0$ &  $(\epsilon  \Upsilon_0)^2/3\sqrt{2}$  & $\epsilon^2\Upsilon_0/a_0^2$ \\ 
$ii$ & $ > 1$ & $< 1/2$  & $1/\beti\ \ \quad (> 1)$ &
$2 a_0 \beti / \epsilon $ & $ \epsilon^2/ 6 \sqrt{2} \Upsilon_0
\beti^3$ & $\epsilon^2/ 4 \Upsilon_0 \beti^2 a_0^2$ \\ 
$iii$ & $< 1$  & $>1/2$ & $2 \Upsilon_0 \bete \beti\ \ (> 1)$ &
$a_0/ \epsilon \Upsilon_0 \beti \bete $ & $(\epsilon \Upsilon_0 \bete
\beti)^2/3\sqrt{2}$ & $\epsilon^2 \Upsilon_0 \bete \beti /a_0^2$ \\
$iv$ & $< 1$ & $< 1/2$  &  $\bete \qquad (> 1)$  &
$2 a_0/ \epsilon \bete $ & $(\epsilon \bete)^2/ 6 \sqrt{2}\Upsilon_0 \beti$ &
$\epsilon^2 \bete /4 \Upsilon_0 \beti a_0^2$ \\ 
\hline
\end{tabular}
\end{center}
\end{table*}

\subsection{Wind launching criterion}
\label{subsec:launching}

We identify the wind launching condition using a similar argument to the
one developed in WK93. The essence of the argument is that, if no wind
were to form, $\wz$ would remain small and
equation~(\ref{eq:z_motiond1}) would be valid throughout the vertical
extent of the disc. The right-hand side of this expression is $<0$
near the midplane, but if it became $>0$ at some height, so that
\begin{equation}
\frac{a_0^2}{\rhot} 
(\jr \ba -\jf \br) > \zt \, ,
\label{eq:launch1}
\end{equation}
this would represent an unphysical situation and would lead to a
neccessary condition for wind launching. We obtain this condition by
differentiating both sides of equation~(\ref{eq:launch1}) with respect
to $\zt$ after substituting for $\jf$ and $\jr$ from
equation~(\ref{eq1:matrix}). In the upper layers of the disc the $\zt$
derivatives of $\rhot$, $\br$ and $\ba$ can be neglected in comparison
with the derivative of $w_{{\rm E}r}$ (given by
equation~\ref{eq:phi_inductiond}). The wind-launching condition then
becomes (setting again $\rhot \approx 1$)
\begin{eqnarray}
\lefteqn{\left[3 \psi^2 + \frac{3}{2}\psi \beta - (1 + q\beta^2)\right]
b_{r{\rm b}}^2 > 
(1 + \beta^2) (1 + q^2 \beta^2) + \psi^2}
\nonumber \\
& & \mbox{ } 
+ \frac{5}{2} \beta \psi + (1 + q\beta^2) b_{\phi{\rm b}}^2 + 
\frac{3}{2}\psi q\beta^2 b_{r{\rm b}} b_{\phi{\rm b}} \, .
\label{eq:launch}
\end{eqnarray}
In the ambipolar diffusion regime, where $\beta$, $q \beta$ and $q
\beta^2$ are $\ll 1$, this expression reduces to equation~(4.16) of
WK93. That study moreover showed that viable solutions in this regime
satisfy $\psi \gtrsim 1$ and $b_{r{\rm b}}>|b_{\phi{\rm b}}|$ (see also
Section~\ref{sec:discuss}), from which it follows that this constraint
further reduces to the ideal-MHD wind-launching criterion $b_{r \rm b} >
1/\sqrt{3}$ (BP82) in this case. Using $b_{r{\rm b}} \approx
\sqrt{2}/a_0$ (see equation~\ref{eq:rphi_B}), this condition can be
expressed as $a_0 \lesssim \sqrt{6} \approx 2$. The corresponding
constraints in the Hall regime are contained in the second column of
inequalities in Table~\ref{table:constraints}. Given that the
approximation $b_{r{\rm b}} \approx \sqrt{2}/a_0$ holds in the Hall
domain as well (see Section~\ref{subsec:base}), it is seen that the
ideal-MHD limit of the wind-launching criterion also characterizes
Cases~$(i)$ and~$(iii)$ in this regime. In the other two Hall
sub-regimes (Cases~$ii$ and~$iv$) $b_{r \rm b}$ is required to exceed
$(2/\Upsilon_0\beti)^{1/2}/\sqrt{3}$, which is $>1/\sqrt{3}$ (i.e. the
minimum inclination angle of the surface field to the rotation axis that is
required for launching a wind is higher than the ideal-MHD value of
$30^\circ$).

\subsection{Location of the base of the wind}
\label{subsec:base}

In light of both theoretical and observational arguments
\citep[e.g.][]{KP00}, only a small fraction of the disc material should
participate in the outflow. This condition is implemented by requiring
that the base of the wind (which we identify with the height above which
the azimuthal velocity becomes super-Keplerian) be located above the
magnetically reduced density scale-height (i.e. $\tilde z_{\rm b}/\tilde
h > 1$). When this condition is violated, $B_\phi$ changes sign within
the disc, reflecting the attempt by the field to transfer angular
momentum to matter before the outflow is initiated (see Fig.~6 in WK93
and Paper~II).

To formulate this constraint, we begin by estimating the vertical
location of the base of the wind. In the hydrostatic approximation, the
azimuthal velocity increases to a Keplerian value (i.e. $w_\phi =0$)
when $w_{{\rm E}r}$ decreases to $0$ (see equations~\ref{eq:r_motiond}
and~\ref{eq1:matrix}). By equation~(\ref{eq:phi_inductiond}), this
occurs at
\begin{equation}
{\tilde z_{\rm b}} \approx \frac{2}{3} \frac{w_{{\rm E}r0}}{b_{r{\rm b}}} \ .
\label{eq:zb}
\end{equation}
Next, we substitute $j_{r0}$ from equation (\ref{eq:phi_alg}) and
$w_{r0} = - \epsilon$ into equation (\ref{eq1:matrix}) to get
\begin{equation}
w_{{\rm E}r0} = \frac{\epsilon}{2 \psi}\ \frac{\psi^2 + (5/2) \psi
\beta +  (1+\beta^2)(1+q^2\beta^2)}{1 + q \beta^2} \ ,
\label{eq:er}
\end{equation}
where, in the Hall regime ($\beta^2 \gg 1$ and $q^2\beta^2 \ll 1$), the
product $(1+\beta^2)(1+q^2\beta^2)$ on the right-hand side reduces to
$\beta^2$.\footnote{A more general expression for $w_{{\rm E}r0}$, in
which no restriction is placed on the value of $q$, is given by
equation~(A11) in WK93.}  As an aside, we note that
equation~(\ref{eq:er}) can be used to derive a parameter consistency
constraint based on the requirement that $w_{{\rm E}r0}$ (or,
equivalently, the midplane value of $K$; see equation~
\ref{eq:K}) must be
$>0$ (see Section~\ref{subsec:dimensionless}). Indeed, treating this
expression as a quadratic equation for $\beta$ and setting it equal to
$0$, we find that viable solutions in the Hall regime require $\beta < -
2 \psi$ or $\beta > - \psi/2$. In view of equation~(\ref{eq:beta}), it
follows that
\begin{equation}
\beta > - \psi/2  \qquad {\rm in\ Hall\ regime.}
\label{beta_eta}
\end{equation}

In the wind-driving discs that are of interest to us the density
scale-height ($\tilde h$) is generally reduced by magnetic squeezing
($\tilde{h} \equiv h/h_{\rm T} < 1$; see Section~\ref{sec:discuss} and
WK93). To obtain an expression for $\tilde h$ we first approximate the
surface values of $\br$ and $\ba$ by $b_{r{\rm b}} \approx
\tilde{z}_{\rm b} (db_r/d\zt)_0 $ and $b_{\phi{\rm b}} \approx
\tilde{z}_{\rm b} (d\ba/d\zt)_0$, respectively.\footnote{We will not use
these expressions to estimate the actual magnitudes of the transverse
magnetic field components at the base of the wind (rather than just
their ratio) both because the assumption of a linear scaling of the
field components with $\zt$ is not expected to remain accurate above a
density scale height and because of the approximation involved in the
definition~(\ref{eq:zb}) of $\tilde{z}_{\rm b}$; instead, we will use
equation~(\ref{eq:rphi_B}) for this purpose.} Then, using
equation~(\ref{eq:B_rB_phi}), we deduce
\begin{equation}
\left | \frac{b_{r{\rm b}}}{b_{\phi{\rm b}}}  
\right |
\approx  \left| \frac{db_r}{db_{\phi}} \right |_0 = \frac{2 \psi +
\beta}{1 + q\beta^2} \ ,
\label{eq:B_rbB_phib}
\end{equation}
which is readily verified to be $>1$ in all the Hall sub-regimes. 
These results are presented in
column 4 of Table~\ref{table:constraints1}, which summarizes
the key properties of our wind solutions for the various Hall
sub-regimes considered in this section. Neglecting the $\ba$
term in equation~(\ref{eq:z_motiond1}) and approximating 
$\rhot \approx 1$ and $\br \approx \zt(d\br/d\zt)_0$, we find
 \begin{equation}
 \frac{d\rhot}{d\zt} \approx - \left[a_0^2 \left(\frac{d\br}{d\zt}
 \right)_0^2 +1   \right] \zt \ . 
 \label{eq:squeez}
 \end{equation}
As we show in Section~\ref{sec:discuss}, magnetic compression
(represented by the first term on the right-hand side of
equation~\ref{eq:squeez}) dominates tidal squeezing (the second term on
the right-hand side) in all viable wind-driving disc solutions. 
Defining the scale-height $\tilde h$ by relating it to an effective
Gaussian density distribution, so that
\begin{equation}
\frac{d\rhot}{d\zt} = -\frac{\zt}{\tilde{h}^2}\ ,
\label{eq:h}
\end{equation}
we thus infer
\begin{equation}
\tilde h \approx \frac{1}{a_0} \left
(\frac{d\br}{d\zt}\right)_0^{-1}\, .
\label{eq:ht}
\end{equation}

By substituting for $(d\br/d\zt)_0$ in equation~(\ref{eq:ht}) from
equations (\ref{eq:phi_ampered}) and (\ref{eq1:matrix}), we can
express $\tilde h$ in terms of the disc parameters as
\begin{equation}
\tilde h \approx \frac{2  a_0}{\epsilon}\ \frac{1 + q
\beta^2}{2\psi + \beta} \ .
 \label{eq:ht1}
\end{equation}
Similarly, combining equations~(\ref{eq:rphi_B}), (\ref{eq:zb}), (\ref{eq:er})
and~(\ref{eq:ht1}) we get the desired expression for the location of the
base of the wind in the Hall regime:
 \begin{equation}
\frac{{\tilde z}_{\rm b}}{\tilde h} \approx \frac{\epsilon^2 (2 \psi + 
\beta)}{6 \sqrt{2} \psi}\  \frac{\psi^2 + (5/2) \psi \beta + \beta^2}{(1 + q
\beta^2)^2} \ .
 \label{eq:zm_ht}
 \end{equation}

Expressions for $\tilde{h}$ valid for Cases $(i) - (iv)$ are listed in
column 5 of Table~\ref{table:constraints1}. Equation~(\ref{eq:zm_ht}),
which is listed in column 6 of that table, can be used to express the
constraint ${\tilde z}_{\rm b}/\tilde h > 1$ in each of the four Hall
sub-regimes. This constraint is shown in the third column of
inequalities in Table~\ref{table:constraints}.
 
\subsection{Dissipation rate}
\label{subsec:dissip}

This constraint, represented by the last set of inequalities in
Table~\ref{table:constraints}, limits the rate of heating by Joule
dissipation at the midplane to less than the rate of gravitational
potential energy released at that location \citep{Kon97},
\begin{equation}
(\j \bdot {\bmath e}^\prime)_0 < \frac{\epsilon}{2 a_0^2}
\frac{\vk}{c_{\rm s}}\ ,
\label{eq:joule}
\end{equation}

\noindent
where, by equations~(\ref{eq1:E-J2}), (\ref{eq:sigma_e}) and~(\ref{eq:j_z}),
\begin{eqnarray}
(\j \bdot {\bmath e}^\prime)_0 &=& \frac{a_0^2}{\psi} (1 + q\beta^2)
(j_{r0}^2 + j_{\phi 0}^2) \cr
&=& \frac{\epsilon^2}{4\psi a_0^2}(1+q\beta^2)\left[ 1 + \left (
\frac{2\psi+\beta}{1+q\beta^2}\right )^2 \right ]\ .
\label{eq:OhmAD}
\end{eqnarray}
The expression~(\ref{eq:OhmAD}) is shown in column 7 of
Table~\ref{table:constraints1} for each of the Hall sub-regimes.

\section{The Ohm Regime}
\label{sec:Ohm_constraints}

In this section we consider the parameter constraints on viable
wind-driving disc solutions for discs in the Ohm diffusivity regime. The
corresponding constraints for discs in the Hall regime were derived in
Section~\ref{sec:constraints} using the hydrostatic approximation, and
here we either directly apply the results presented in that section or
else further generalize them to cover also the Ohm domain. We again adopt
the multi-fluid formalism in a weakly ionized ``ion''-``electron'' gas
with $q \ll 1$, in which the Ohm regime is defined by the inequalities
$|\beti| \ll |\bete| \ll 1$ (or, equivalently, $|\beta| \gg q|\beta| \gg
1$). We also continue to assume that $\epsilon_{\rm B} \approx 0$.

\subsection{Sub-Keplerian flow below the launching region}
\label{subsec:Ohm_subK}

All the results derived in Section~\ref{subsec:subK} are applicable in
this regime. In particular, a necessary condition for a solution to
exist is that the inequality $a_0 >C^{-1/2}$ is satisfied, where $C$,
given by equation~(\ref{eq:rphi_B1}), is equal to $2\psi$ (where $\psi =
\Upsilon_0$) or to $\beta$ in the limits where $\psi/|\beta|$ is $\gg
1/2$ or $\ll 1/2$, respectively. The magnitude of this parameter
combination thus provides a natural classification criterion, similarly
to the situation in the Hall regime. However, in contrast with the Hall
case, in which the factor $q\beta^2$ in the denominator of
equation~(\ref{eq:B_rB_phi}) can be either $>1$ or $<1$, in the Ohm
limit $q\beta^2 \gg 1$ and this parameter combination cannot be used to
identify the relevant parameter sub-regimes. A suitable combination can
nevertheless be found by using the relation~(\ref{eq:B_rB_phi}) to
determine which transverse magnetic field component (radial or
azimuthal) dominates at the disc surface (see
equation~\ref{eq:B_rbB_phib}). In the limit $|\beta| \gg 2\psi$ we have
$|b_{r{\rm b}}/b_{\phi{\rm b}}| = 1/q\beta \ll 1$, but in the opposite
limit ($2\psi\gg |\beta|$) this ratio is given by $|b_{r{\rm
b}}/b_{\phi{\rm b}}| = 2\psi/q\beta^2$, which can be either $> 1$ or $<
1$. This suggests that the parameter combination $\psi/q\beta^2$ could
serve to sub-divide the parameter regime $\psi/\beta > 1/2$. As we
verify in the subsequent subsections, this is indeed a proper
choice. This result is not surprising given the fact that, in the Ohm
regime, $\psi/q\beta^2$ is equal to the Elsasser number $\Lambda_0$
(defined in Section~\ref{subsec:param}) and that $\Lambda_0$ plays a
similar role in the Hall regime (where it is equal to $\psi/|\beta|$).

\subsection{Wind launching criterion}
\label{subsec:Ohm_launching}

In analyzing the inequality given by equation~(\ref{eq:launch}) in the
Ohm limit, we consider separately the cases $\psi/|\beta| > 1/2$ and
$\psi/|\beta| < 1/2$. As we noted in Section~\ref{subsec:Ohm_subK}, the
first case can be sub-divided according to whether $\Lambda_0$ is $ >
1/2$ or $< 1/2$. When $\Lambda_0 > 1/2$, the dominant terms on both the
left-hand side and the right-hand side of equation~(\ref{eq:launch}) are
the ones containing the factor $\psi^2$, and we recover the BP82
launching condition of a centrifugally driven, ideal-MHD wind from a Keplerian
disc, $b_{r{\rm b}} > 1/\sqrt{3}$. However, the application of the above
equation to the second sub-regime, in which the inequalities $|\beta|
<2\psi < q\beta^2$ are satisfied, is less straightforward. Indeed, while
one can readily verify that the term involving $\psi^2$ again dominates
the left-hand side of equation~(\ref{eq:launch}) in this case, on the
right-hand side of this equation either the first term or the last term
could potentially dominate. Assuming that the first term ($\sim
q^2\beta^4$) in fact dominates, one can substitute $b_{r{\rm b}}^2
\approx (2/a_0^2)(2\psi/q\beta^2)^2$ from equations~(\ref{eq:rphi_B})
and~(\ref{eq:B_rbB_phib}) into equation~(\ref{eq:launch}) to infer that
the factor $6\psi^2/a_0^2q^2\beta^4$ is $>(q\beta^2/2\psi)^2 >1$. But in
the chosen parameter regime this factor is equal to the ratio of the
absolute magnitude of the last term on the right-hand side of
equation~(\ref{eq:launch}) to the first term on the right-hand side of
this equation, which is a contradiction (as we assumed that the first
term on the right-hand side dominates). A self-consistent solution can
thus be obtained only if the ratio of the absolute values of the last
and first terms on the right-hand side of equation~(\ref{eq:launch}) is
required from the start to exceed 1, in which case the inequality
expressed by this equation would be trivially satisfied (with the
left-hand side being $>0$ and the right-hand side -- which is
proportional to $b_{r{\rm b}}b_{\phi{\rm b}}$ -- being $<0$). This
requirement, in turn, provides an effective wind-launching condition on the
inclination of surface poloidal field: $b_{r{\rm b}}>
2/\sqrt{3}$.\footnote{This condition implies a minimum inclination angle
of the surface poloidal field to the rotation axis of $\simeq 49^\circ$, as
compared with $30^\circ$ for the ideal MHD-like case ($b_{r{\rm b}} >
1/\sqrt{3}$).}

Turning now to the case $\psi/|\beta| < 1/2$, we find that it is again
possible to choose between two alternatives: either $\psi > q|\beta|$ or
$\psi < q|\beta|$. The first alternative is similar to the just-discussed
$\psi/|\beta| > 1/2$, $\Lambda_0 < 1/2$ sub-regime in that one can
immediately identify the dominant term on the left-hand side of
equation~(\ref{eq:launch}) (which in this instance involves the factor
$3\psi\beta/2$) even as the choice between the first and last terms on
the right-hand side of this equation appears to be unclear. The
assumption that the first term on the right-hand side dominates again
leads to a contradiction, so in this case, too, one is led to impose the
requirement that the last term always dominates the first term on the
right-hand side of equation~(\ref{eq:launch}). This requirement, in
turn, again implies an effective wind-launching criterion, which in this
case is $b_{r{\rm b}}>[(4/3)(\beta/2\psi)]^{1/2}$ ($ >
2/\sqrt{3}$).\footnote{We have implicitly assumed that the Hall
parameters are $>0$ in this case; this is explicitly justified in
Section~\ref{sec:discuss}.} The second alternative, corresponding to the
parameter regime $q|\beta|<\psi<|\beta|/2$, does not allow the launching
condition to be satisfied since it implies that, while the right-hand
side of equation~(\ref{eq:launch}) remains $>0$, the left-hand side is
dominated by the last term and is therefore $< 0$. We therefore exclude
this parameter regime from further consideration in the ensuing
subsections, in which we continue to verify that the three other Ohm
sub-regimes that we have identified are compatible with wind-driving
disc solutions that obey all the relevant physical requirements.

\subsection{Location of the base of the wind}
\label{subsec:Ohm_base}

In the Hall regime we deduced that all viable solutions satisfy
$|b_{r{\rm b}}/b_{\phi{\rm b}}| \approx |db_r/db_{\phi}|_0 \gg 1$,
and we used this approximation in deriving the expressions~(\ref{eq:ht})
and~(\ref{eq:zm_ht}) for the normalized density scale height $\tilde{h}$
and disc height $\tilde{z}_{\rm b}$. However, as already noted in
Section~\ref{subsec:Ohm_subK}, in the Ohm case this inequality could in
principle be reversed, in which case the above expressions would need to
be modified. This can be done through a straightforward generalization
of the derivations presented in Section~\ref{subsec:base}. Using
equation~(\ref{eq:B_rbB_phib}) and defining
\begin{equation}
D(\psi,\beta) \equiv \left ( \frac{d\br}{d\ba}\right )_0^2 = \left
(\frac{2 \psi + \beta}{1 + q\beta^2}\right )^2 \ ,
\label{eq:Ohm_D}
\end{equation}
we obtain
\begin{equation}
\tilde{h} = \frac{2a_0}{\epsilon}  [ 1 + D(\psi,\beta)]^{-1/2}\; ,
\label{eq:Ohm_ht}
\end{equation}
\begin{equation}
b_{r{\rm b}} \approx \frac{\sqrt{2}}{a_0} [1+1/D(\psi,\beta)]^{-1/2}
\label{eq:Ohm_brb}
\end{equation}
and
\begin{equation}
\frac{{\tilde z}_{\rm b}}{\tilde h} \approx \frac{\epsilon^2}{6 \sqrt{2}
\psi}\  \frac{\psi^2 + (5/2) \psi \beta +
(1+\beta^2)(1+q^2\beta^2)}{2\psi + \beta} [1 + D(\psi,\beta)]\, .
 \label{eq:Ohm_zm_ht}
 \end{equation}
It can be readily verified that in the Hall limit these results reduce to 
the corresponding expressions presented in Section~\ref{subsec:base}.

\subsection{Dissipation rate}
\label{subsec:Ohm_dissip}

The results of Section~\ref{subsec:dissip}, given by
equations~(\ref{eq:joule}) and~(\ref{eq:OhmAD}), continue to apply also
in the Ohm regime. Note that the second of the above equations can be
written in the form 
\begin{equation}
(\j \bdot {\bmath e}^\prime)_0 = \frac{\epsilon^2}{4\psi
a_0^2}(1+q\beta^2)\left[ 1 + D(\psi,\beta) \right ] \; . 
\label{eq:Ohm_OhmAD}
\end{equation}

\subsection{Results}
\label{subsec:Ohm_results}

The results of the foregoing analysis are collected in
Tables~\ref{table:Ohm_constraints} and~\ref{table:Ohm_constraints1},
which are patterned on Tables~\ref{table:constraints}
and~\ref{table:constraints1}, respectively, in
Section~\ref{sec:constraints} (where the corresponding data for the Hall
regime are presented). The implications of these results are discussed
in conjunction with those of our findings for the Hall and
ambipolar-diffusion regimes in Section~\ref{sec:discuss}.

\setlength{\cellspacetoplimit}{1mm}
\setlength{\cellspacebottomlimit}{1mm}
\begin{table*}
\caption{Parameter constraints for wind-driving disc solutions in the
limit where the Ohm diffusivity dominates. Three distinct cases can be
identified, as indicated. The meaning of the four inequalities in each
sub-regime is the same as in Table~\ref{table:constraints}.
The parameters $\psi$, $\beta$ and $q\beta$ used in the text are related
to the parameters employed in the table through $\psi = \Upsilon_0$,
$\beta = 1/\beti$ and $q\beta = 1/\bete$.
}
\label{table:Ohm_constraints}
\begin{center}
\begin{tabular}{Sc|Sc|Sc|ScScScScScScScScSc}
\hline
{\normalsize Case} & \multicolumn{2}{c}{{\normalsize Limits}} &
\multicolumn{9}{c}{{\normalsize Parameter Constraints -- Ohm Limit }} \\
& $\Upsilon_0 |\beti|$ &$\Lambda_0=\Upsilon_0 |\bete| |\beti|$ & 
 \multicolumn{9}{c}{{\small (multi-fluid formulation)}} \\
\hline 
$i$ &
$> 1/2$ & $> 1/2$ &
$(2\Upsilon_0)^{-1/2}$ &  $\lesssim$ & $a_0$ & $\lesssim$ & $2$ &
$\lesssim$ & $\epsilon\Upsilon_0\bete \beti$ &
$\lesssim$ & $\vk/2 c_{\rm s}$   \\
$ii$ & $> 1/2$ & $< 1/2$ & 
$(2\Upsilon_0)^{-1/2}$ & $\lesssim$ & $a_0$ &  $\lesssim$ & $2\Upsilon_0\bete \beti$
& $\lesssim$ & $\epsilon/2$ &  $\lesssim$ & $\Upsilon_0 \bete \beti \vk/
c_{\rm s}$\\
$iii$ & $< 1/2$ & $> |\beti|$ & 
$\beti^{1/2}$ & $ \lesssim$ &  $a_0$ & $\lesssim$ &  $2 (\Upsilon_0
\beti)^{1/2} \bete$ &   $\lesssim$ & $\epsilon/2$ & 
$\lesssim$ & $\Upsilon_0 \bete \beti \vk/c_{\rm s}$ \\
\hline
\end{tabular}
\end{center}
\end{table*}

\setlength{\cellspacetoplimit}{1mm}
\setlength{\cellspacebottomlimit}{1mm}
\begin{table*}

\caption{Key properties of viable disc solutions in the Ohm
regime. See the caption to Table~\ref{table:constraints1} for a
description of the listed quantities.
}
\label{table:Ohm_constraints1}
\begin{center}
\begin{tabular}{Sc|Sc|Sc|SrScScSc}
\hline
{\normalsize Case} & \multicolumn{2}{c}{{\normalsize Limits}} &
\multicolumn{4}{c}{{\normalsize Solution Characteristics -- Ohm Limit }} \\
& $\Upsilon_0 |\beti|$ & $\Lambda_0=\Upsilon_0 |\bete| |\beti|$&
$|db_r/db_\phi|_0$ &  $\tilde
h$ & $\tilde z_{\rm b}/\tilde h$ &  
$(\j \bdot {\bmath e}^\prime)_0 $ \\ 
\hline 
$i$ & $> 1/2$ & $> 1/2$ & 
$2 \Upsilon_0 \bete \beti\ \ (> 1)$ & $a_0/ \epsilon \Upsilon_0 \bete \beti $ &
$(\epsilon \Upsilon_0)^2/3\sqrt{2}$  & $\epsilon^2\Upsilon_0 \bete \beti /a_0^2$ \\ 
$ii$ & $>1/2$ & $< 1/2$ &
$2 \Upsilon_0 \bete \beti\ \ (< 1)$ & $2 a_0/ \epsilon $ &
$(\epsilon /\Upsilon_0 \bete \beti)^2/12\sqrt{2}$ & $\epsilon^2/4\Upsilon_0 \bete \beti
a_0^2$ \\ 
$iii$ & $< 1/2$ & $ > |\beti|$ & $\bete \qquad (< 1)$ &
$2 a_0/ \epsilon $ & $ (\epsilon/\bete)^2/ 6 \sqrt{2} \Upsilon_0
\beti$ & $\epsilon^2/4\Upsilon_0 \bete \beti a_0^2$ \\ 
\hline
\end{tabular}
\end{center}
\end{table*}

\subsection{Formulation in terms of scalar conductivity}
\label{subsec:Ohm_conductivity}

As we noted in Section~\ref{subsec:model}, the analysis of the problem
in the Ohm regime can be alternatively carried out in terms of the scalar
conductivity $\sigma_{\rm O}$ (equation~\ref{eq:sig1}). In the case of
a two-component plasma with $q \ll 1$ this is just the ``electron'' electrical
conductivity $\sigma_{\rm e}$ (equation~\ref{eq:sigma_e}). With our
adopted normalization (see equation~\ref{eq:nondim3}), the corresponding
dimensionless conductivity at the midplane of the disc is
\begin{equation}
\tilde{\sigma}_{\rm e 0} = \frac{4 \pi h_{\rm T} c_{\rm s}\sigma_{\rm
e}}{c^2} = \frac{h_{\rm T} c_{\rm s}}{\eta_{\rm O 0 }} = \frac{\Upsilon_0 \bete
\beti}{a_0^2}\ .
\label{eq:Ohm_sigma_e}
\end{equation}

From equation~(\ref{eq:Ohm_sigma_e}) it is seen that, in the Ohm regime,
$\Lambda_0 = a_0^2 \tilde{\sigma}_{\rm e}$.\footnote{This also follows
directly from the relation~(\ref{eq:bi}) in this limit.} This makes it
possible to relate the results derived in this section to the equivalent
scalar-conductivity formulation. In particular, Cases~($i$) and~($ii$)
identified above correspond to $a_0^2 \tilde{\sigma}_{\rm e}$ being
$>1/2$ and $<1/2$, respectively. Resistive-MHD disc models that have
appeared in the literature typically use Ohm's law in the form $\E = -
{\v \cross \B}/{c} + {\J}/{\sigma_{\rm e}}$ and ignore the second (Hall)
term on the right-hand side of the more general expression given by
equation~(\ref{eq1:E-J}). On the face of it, this might be justified by
the fact that the ratio of the conductivity prefactors of the second and
first terms on the right-hand side of this equation, $|\sigma_{\rm H}|
\sigma_{\rm O}/\sigma_\perp^2$, is $\sim |\beta_{\rm e}|$, which is $\ll
1$ in the Ohm regime. However, this argument does not account for the
fact that the Hall term in the expression for $\E'$ is perpendicular to
$\J$, which implies, in particular, that the Hall term appears in the
prefactor of the {\em radial} current density $J_r$ in the normalized
expression for the {\em azimuthal} electric field component $E_\phi$
(equation~\ref{eq:epsilonB}). The only other contributor to this
prefactor at $z=0$ is associated with the advective ($\propto {\v \cross
\B}/{c}$) term, which could in principle be subordinate to the
Hall-current term. Now, these two terms (Hall and advective) in
equation~(\ref{eq:epsilonB}) correspond to the two terms ($\beta$ and
$2\psi$, respectively) in the numerator of the expression for
$|d\br/d\ba|_0$ (equation~\ref{eq:B_rbB_phib}) that are used in our
solution classification scheme for the Ohm regime. In Cases~($i$)
and~($ii$) $\Upsilon_0 |\beti|= a_0^2 \tilde {\sigma}_{\rm e 0}/|\bete|
> 1/2$ and so the Hall term can be neglected, but in Case~($iii$) this
inequality is reversed and the Hall term dominates.

\section{Discussion}
\label{sec:discuss}

The first term within the first pair of parentheses on the right-hand
side of equation~(\ref{eq:OhmAD}) represents the ohmic contribution to
the Joule dissipation in the disc, whereas the second term corresponds
to the ambipolar dissipation. (The Hall term in Ohm's law, i.e., the
term $\propto \J \cross \B$ in equations~\ref{eq1:E-J}
and~\ref{eq1:E-J2}, does not contribute to the $\J \bdot \E^\prime$
dissipation.) This suggests \citep[see][]{Kon97} that the slip factor
$s_0=1/q\beta^2=\bete\beti$ can be used to distinguish between ambipolar
diffusion-like ($\bete\beti > 1$) and Ohm-like ($\bete\beti < 1$)
sub-regimes in the nominal Hall parameter domain ($|\beti| < 1 <
|\bete|$).\footnote{This can also be seen directly from
equation~(\ref{eq:etaA1}), which implies that $s = \etaA/\etaO$.}
For comparison, both $\bete\beti > 1$ {\em and} $|\beti| > 1$
must hold in the actual ambipolar diffusion regime, whereas both
$\bete\beti < 1$ {\em and} $|\bete| < 1$ must be satisfied in the
genuine Ohm regime.  This criterion forms the basis of the
classification given in the second column of
Table~\ref{table:constraints}. The relevance of this criterion is
demonstrated by the fact that the first row of inequalities (Case~$i$)
exactly reproduces the corresponding constraints in the ambipolar
diffusion case (see WK93 and \citealt{Kon97}) and that the third row
(Case~$iii$) is identical to the first row of inequalities in
Table~\ref{table:Ohm_constraints} (Case~$i$
in the Ohm regime).\footnote{These correspondences apply also to the
solution characteristics listed in Tables~\ref{table:constraints1}
and~\ref{table:Ohm_constraints1}.}

\begin{figure}
\centering
\includegraphics[width=0.45\textwidth]{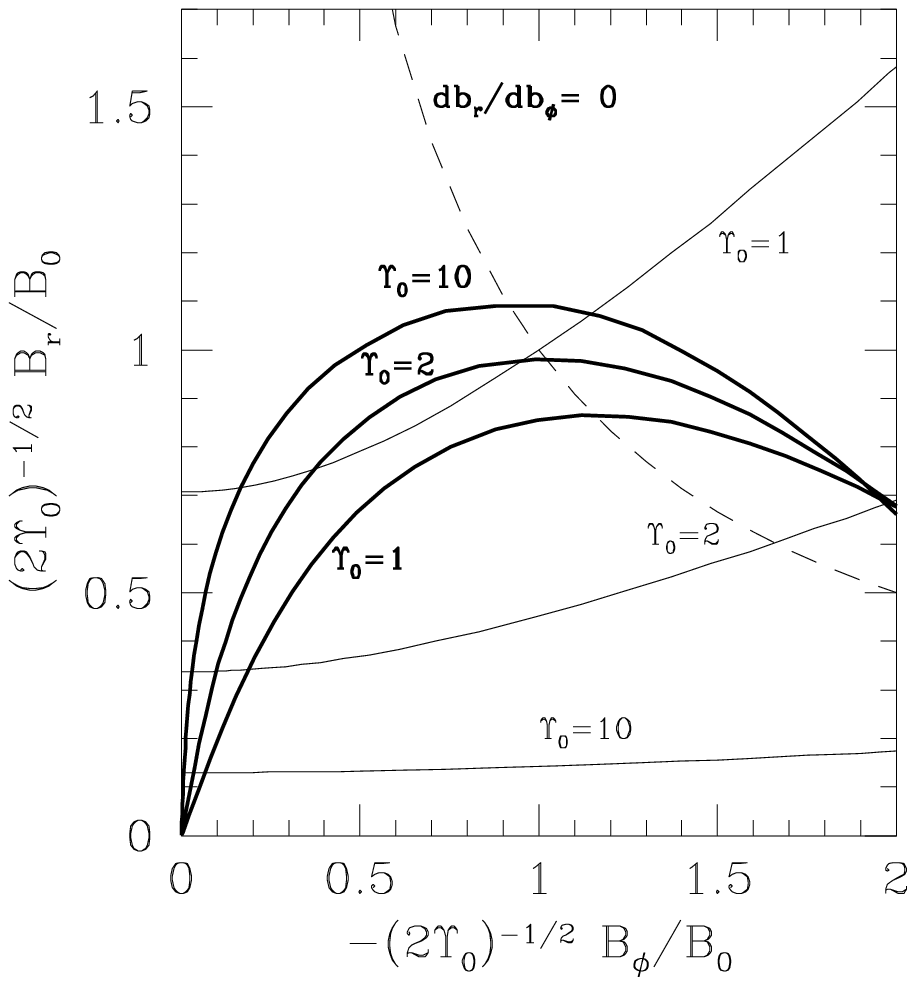} 
\caption{Plots of $b_r$ versus $b_{\phi}$ (each normalized by
$(2\Upsilon_0)^{1/2}$) in the quasi-hydrostatic
region of the disc (dark solid lines) for $\Upsilon_0 = 1$, $2$ and $10$ in
the limit where ambipolar diffusion dominates. 
Physically viable solutions
must satisfy two constraints. First, they must
lie to the left of the long-dashed line, which marks the point where $d
b_r/d b_{\phi}$ changes sign: this ensures that the azimuthal
velocity is sub-Keplerian within the disc (i.e. $w_{\phi}$, given by
equation~\ref{eq:wf}, is $< 0$ for $\tilde z < \tilde z_{\rm
b}$). Second, they must also lie above the light solid line for the
corresponding value of $\Upsilon_0$, so that the wind-launching criterion
(equation~\ref{eq:launch}) is satisfied. It is evident from the figure
that these two constraints together imply $\Upsilon_0 \gtrsim 1$.}
\label{fig:AD_data} 
\end{figure}

\begin{figure*}
\centering
\includegraphics[width=0.8\textwidth]{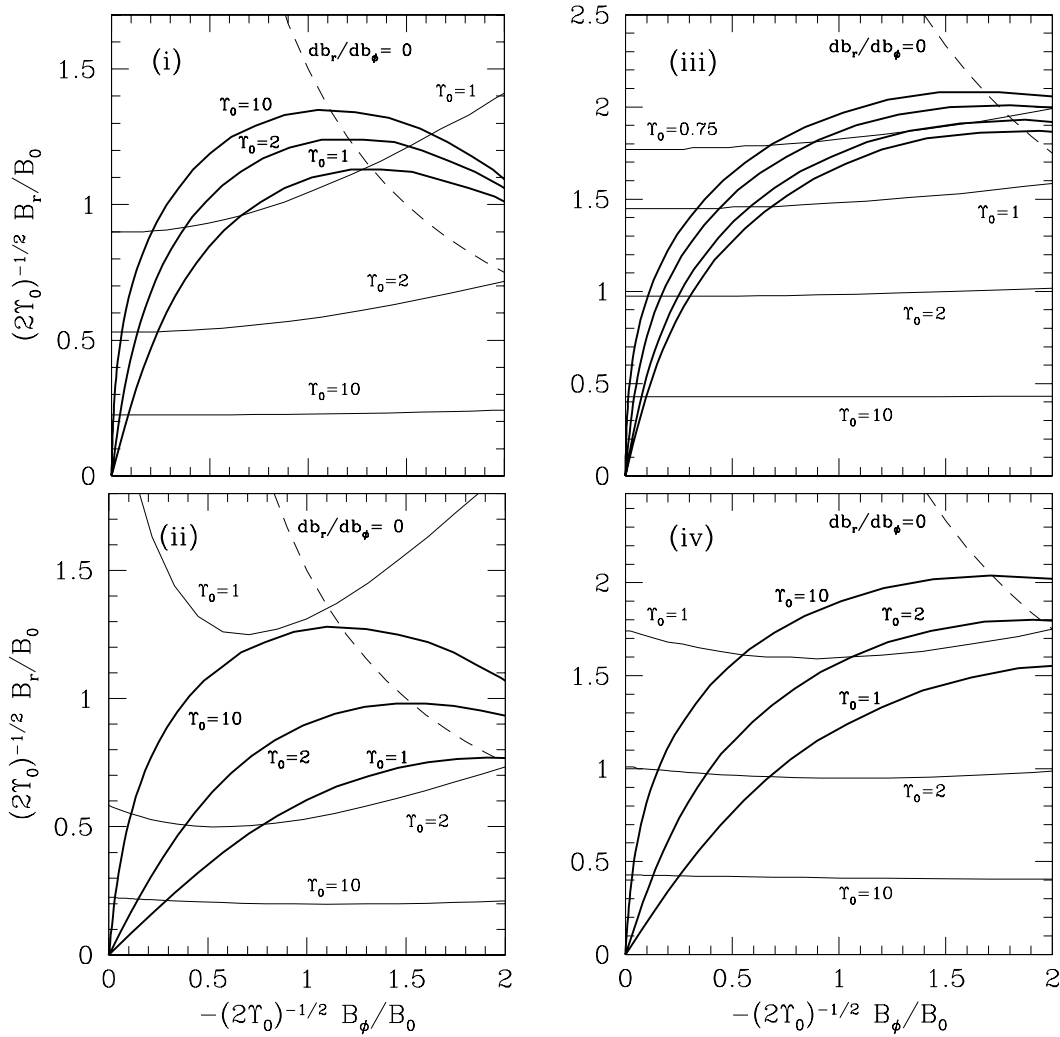}
\caption{Same as Fig.~\ref{fig:AD_data}, but in the limit where the Hall
diffusivity dominates. The four panels correspond to Cases ($i$) through
($iv$) in Table~\ref{table:constraints}. Note that, as in the
ambipolar diffusion regime shown in Fig.~\ref{fig:AD_data}, physically
viable solutions satisfy $\Upsilon_0 \gtrsim 1$ also in each of the Hall
sub-regimes.} 
\label{fig:Hall_data}
\end{figure*}

The third column in Table~\ref{table:constraints} classifies the
solutions according to the value of the magnetic coupling parameter
$\Lambda_0$ (the Elsasser number, which in the Hall regime is equal to
$\Upsilon_0 |\beti|$). Although the constraints obtained for large ($\Lambda_0
> 1/2$; Cases~$i$ and~$iii$) and small ($\Lambda_0 < 1/2$; Cases~$ii$
and $iv$) values of $\Lambda_0$ are clearly distinct in their forms, it
is interesting to note that the sub-Keplerian rotation and
wind-launching requirements (the first and second inequalities,
respectively) together imply that $\Upsilon_0 \gtrsim 1$ {\em in all
cases}. This inequality already follows directly from the conditions
$\Upsilon_0 |\beti| > 1/2$ and $|\beti| < 1$ that define the Hall Cases~$(i)$
and~$(iii)$, but it is noteworthy that all viable solutions in the Hall
domain satisfy this constraint, which was previously inferred in the
ambipolar diffusion regime (see WK93). This result can be
visualized by plotting the sub-Keplerian-rotation and wind-launching
constraints for each of the Hall sub-regimes. First, however, we present
the relevant curves in the ambipolar diffusion limit
(Fig.~\ref{fig:AD_data}; cf. Fig.~3 in WK93). The dark, solid lines in
this figure show the run of $b_r$ versus $b_{\phi}$, computed from
equation~(\ref{eq:B_rB_phi}), for $\Upsilon_0 = 10$, $2$ and $1$. In fully
sub-Keplerian discs $d b_r/ d b_{\phi}$ must be $< 0$ below the base of
the wind (i.e. for $\tilde z < \tilde z_{\rm b}$; see
equations~\ref{eq:wf} and~\ref{eq:B_rB_phi}). This condition
constrains viable solutions to lie {\em to the left} of the long-dashed
curve that marks the locus of points where $db_r/db_{\phi}=0$. On the
other hand, these solutions must also satisfy the wind-launching
criterion (equation \ref{eq:launch}), which constrains them to reside
{\em above} the light solid line for the given value of $\Upsilon_0$. It is
evident from these plots that no physically viable solutions exist for
$\Upsilon_0 < 1$. The corresponding curves for Cases~$(i)-(iv)$ in the Hall
limit are shown in the four panels of Fig.~\ref{fig:Hall_data}: they
confirm that viable solutions satisfy $\Upsilon_0 \gtrsim 1$ also in all
the Hall sub-regimes.

As was shown in Section~\ref{sec:Ohm_constraints}, the magnitude of
the parameter combination $\Upsilon_0 |\beti|$ compared to $1/2$ is
one of the classification criteria also in the Ohm regime (see the
second column in Table~\ref{table:Ohm_constraints}). When $\Upsilon_0
|\beti| > 1/2$ the second classification criterion in the Ohm regime
can again be expressed in terms of the Elsasser number $\Lambda_0$
(which in this limit equals $\Upsilon_0 \bete \beti$) being either
$>1/2$ (the Ohm Case~$i$) or $<1/2$ (the Ohm Case~$ii$; see the third
column in Table~\ref{table:Ohm_constraints}). Similarly to the
situation in the Hall Cases~($i$) and~$(iii)$, the fact that
$|\beti|<1$ implies that $\Upsilon_0$ must be $\gtrsim 1$ also in
these two Ohm sub-regimes. The third Ohm sub-regime is defined by
$\Upsilon_0|\beti|< 1/2$ and $\Upsilon_0|\bete|> 1$. Since $|\bete| <
1$ in the Ohm domain, it follows that $\Upsilon_0 \gtrsim 1$ in this
case too.

The parameter $\Upsilon_0$ (which is equal to $\Lambda_0$ in the
ambipolar diffusion regime) thus turns out to be of fundamental
importance in the theory of diffusive wind-driving discs in that it is
inferred to be $\gtrsim 1$ for all viable solutions irrespective of the
conductivity regime. Physically, the condition $\Upsilon_0 \gtrsim 1$
expresses the requirement that the momentum exchange time of the
neutrals with the particles that dominate the momentum of the
charged species (the comparatively massive ``ions'' in our formulation)
be shorter than the dynamical time (i.e. the Keplerian orbital
time). This requirement is evidently more basic than having the nominal
neutral--field coupling parameter $\Lambda_0$ be $\gtrsim 1$. In fact,
as noted above, some of the distinct parameter sub-regimes that we
identified are defined by the inequality $\Lambda_0 < 1/2$. However,
even in the latter cases $\Lambda_0$ remains bounded from below by a
numerical factor that typically is not $\ll 1$ ($\Lambda_0 \gtrsim
a_0^2/3$ and $\Lambda_0 \gtrsim a_0/\sqrt{6}$ in the Hall and Ohm
regimes, respectively, from the wind launching condition; see
Tables~\ref{table:constraints} and~\ref{table:Ohm_constraints}). 

The lower bounds on $\Lambda_0$ reflect our assumption that the
modelled discs are strongly coupled, which implies, in particular, that
the transverse magnetic field component starts to grow already at $\zt =
0$ (see the solution curves presented in Paper~II). As was first
demonstrated by \citet{Li96} and \citet{War97}, viable solutions of
discs in which the bulk of the matter is not well coupled to the field
can also be constructed.\footnote{Such configurations were later found
to arise naturally in certain models of disc formation from the collapse
of rotating molecular cloud cores \citep[see][]{KK02}.} In these
solutions, significant field-line bending that enables a centrifugally
driven wind to be launched from the disc surface only starts well above
the midplane. Although the basic disc properties above the height where
$\br$ starts to increase rapidly are similar to those of strongly
coupled configurations, this location depends on the detailed density
and ionization structure of the disc and cannot be readily determined a
priori. It roughly coincides with the height where $|d\br/d\ba|$
(estimated using equation~\ref{eq:B_rbB_phib}), which is typically $\ll
1$ at the midplane under these circumstances, has grown to a value $\sim
1$. However, disc material already moves inward below this height on
account of the fact that $|\ba|$ becomes $\lesssim 1$ when $\br$ is
still $\ll 1$, which means that a significant magnetic torque (of
density $\propto B_z dB_\phi/dz$) can be exerted even at lower elevations.

The magnitude of the field-strength parameter $a_0$ is bounded from
below by the sub-Keplerian rotation constraint (the first inequality
in Tables~\ref{table:constraints} and~\ref{table:Ohm_constraints})
and from above by the wind-launching constraint (the second inequality
in Tables~\ref{table:constraints} and~\ref{table:Ohm_constraints}). If
the magnetic field is too weak it will not drive an outflow but
instead mediate a ``two-channel'' flow within the disc
\citep[see][]{SKW07}. On the other hand, if the field is too strong it
will not bend out sufficiently to drive a centrifugal wind. Our
typical solutions are characterized by $a_0 \lesssim 1$. We have found
that the ratio $|b_{r{\rm b}}/b_{\phi{\rm b}}|$ of the radial to
azimuthal magnetic field amplitudes at the disc surface is $> 1$ in
the ambipolar diffusion regime (WK93) as well as in all the Hall
sub-regimes (Section~\ref{subsec:base}). However, in the Ohm regime
this is true only for Case~($i$), whereas in the other two Ohm
sub-regimes $|b_{\phi{\rm b}}|> b_{r{\rm b}}$ (see
Table~\ref{table:Ohm_constraints1}). Nevertheless, even in the latter
two cases $b_{r{\rm b}}$ is required to exceed a certain lower bound
(which is at least $2/\sqrt{3}$; see
Section~\ref{subsec:Ohm_launching}) to satisfy the wind launching
condition. It is also worth noting in this connection that, even if
the magnetic field at the disc surface is not bent strongly enough to
satisfy the wind launching requirement, vertical magnetic
angular-momentum transport along a large-scale field that threads the
disc could potentially still take place by other means, such as
magnetic braking (see footnote~\ref{fn:braking}).

By combining the wind-launching and wind-loading constraints (the
second and third inequalities, respectively, in
Tables~\ref{table:constraints} and~\ref{table:Ohm_constraints}) and
using the expression for $\tilde{h}$ in the fifth column of
Tables~\ref{table:constraints1} and~\ref{table:Ohm_constraints1}, one
finds that $\tilde{h} < 1$, i.e. that the vertical confinement of the
disc is primarily magnetic (due to the vertical gradient of ($B_r^2 +
B_\phi^2)/8\pi$) rather than tidal for all the cases in the Hall and
Ohm regimes. This result, which was previously shown by WK93 to apply
to ambipolar diffusion-dominated discs, is corroborated by the
full solutions presented in Paper~II. Predominantly magnetic
confinement is thus seen to be a generic property of our wind-driving
disc model. The generality of this result was, however, questioned by
\citet{Fer97}, who suggested that it might be a consequence of the
simplification $\rho v_z=$const adopted in our treatment of the mass
conservation relation. In addressing this question it is interesting
to note that the inequality $\tilde{h}< 1$ was inferred in this paper
already on the basis of the hydrostatic approximation, which does not
make use of the continuity equation. On the other hand, our argument
involved an estimate of $\tilde{z}_{\rm b}$, the normalized height of
the disc surface, which corresponds to a disc region where the
hydrostatic approximation may no longer be accurate.  A definitive
resolution of this issue would require using a more elaborate model in
which the above simplification is not made.

The wind-loading and energy dissipation constraints (the third and
fourth inequalities, respectively, in Tables~\ref{table:constraints} and
~\ref{table:Ohm_constraints}) provide lower and upper bounds,
respectively, on the radial velocity parameter $\epsilon$. The lower
bound reflects the fact that the height of the disc surface scales as
$\tilde{h}$ whereas the scale height scales as $1/\epsilon$, so that
$\tilde{z}_{\rm b}/\tilde{h}$, which is required to be $> 1$, scales as
$\epsilon^2$ (see the fifth and sixth columns in
Tables~\ref{table:constraints1} and ~\ref{table:Ohm_constraints1}). The
upper bound is a consequence of the fact that both the electric field
and the current density scale as $\epsilon$ (see the seventh column in
Tables~\ref{table:constraints1} and ~\ref{table:Ohm_constraints1}), so
that the Joule dissipation (which scales as $\epsilon^2$) might exceed
the rate of gravitational energy release ($\propto \epsilon$) if the
inflow speed were too high. The solutions presented in Paper~II have
values of $\epsilon$ in the range $\sim 0.3-1$. With these parameters
one infers a nominal emptying time $\lesssim 10^4\; {\rm yr}$ for a
representative protostellar disc, which is consistent with the mean
duration of the earliest (Class~0) phase of protostellar evolution but
is very much shorter than the total age of accreting protostars.  This
is the essence of the ``short evolution time'' critique of pure-wind
angular-momentum transport models of protostellar discs, which has led
to the argument that alternative angular-momentum transport mechanisms
must dominate in such systems \citep[e.g.][]{Shu08}.\footnote{These
alternative mechanisms potentially include vertical transport along a
large-scale open magnetic field by means other than the steady wind that
has been the focus of our discussion; see footnote~\ref{fn:braking}.}
Note, however, that even if a centrifugally driven wind does not control
the overall evolution of the disc throughout its lifetime, it might
still dominate the angular momentum transport in regions of the disc
from which strong outflows are launched, which may be especially
pertinent during the FU Orionis-type rapid accretion episodes that are
thought to account for most of the mass deposition on to low-mass
protostars \citep[e.g.][]{CHS00}. Furthermore, wind-mediated angular
momentum transport could in principle also contribute in regions where a
radial transport mechanism dominates \citep[see][]{SKW07}. It is also
worth noting that significantly lower mean inflow speeds could be
attained in more realistic, vertically stratified wind-driving disc
models \citep[e.g.][]{KS09}, and in particular in {\em weakly coupled}
discs (see \citealt{Li96} and \citealt{War97}). Further work is,
however, required to determine the extent to which this effect would
increase the predicted disc evolution time.

We have been careful to use the absolute values of $\beti$ and $\bete$
in delineating the various sub-regimes in the Hall and Ohm domains in
view of the fact that our definition (equation~\ref{eq:Hall_parameter})
of the Hall parameters has an explicit dependence on the magnetic field
polarity and thus allows them to have both positive and negative
values. However, in our table entries for the Hall Cases~($ii$)
and~($iv$) and for the Ohm Case~($iii$) we implicitly treated the ion
and electron Hall parameters as \emph{positive} quantities. This can be
readily justified by observing that these cases correspond to $|\beta|$
being $>2\psi$ and taking note of the inequality $\beta > -2\psi$
(equation~\ref{eq:beta}) that all viable solutions must obey. These two
inequalities can be simultaneously satisfied only if $\beta > 0$. In
other words, for these parameter sub-regimes a self-consistent
wind-driving disc solution can be obtained only if the field has a
positive polarity but not if ${\rm sgn}\{B_z\} < 0$. The sensitivity to
the field polarity reflects the explicit dependence on the Hall
conductivity in these cases (see Sections~\ref{subsec:model}
and~\ref{subsec:Ohm_conductivity}) and is a specific prediction of our
model that is checked in the numerical work presented in Paper~II. We
emphasize that our discussion here concerns the nature of certain
sub-regimes, in which \emph{only} solutions with $\beta > 0$ are
predicted to exist. This property is distinct from the dependence of
wind-driving wind solutions on the field polarity in other parameter
sub-regimes in which \emph{both} positive and negative values of $\beta$
are allowed but in which changing the sign of $\beta$ modifies the
behaviour of the solution (see WK93). An example of the latter
dependence is given in Paper~II for the Hall Case~($iii$), which admits
$\beta$ values in the range $-\psi/2<\beta<2\psi$ (see
equation~\ref{beta_eta}).

Our analysis has concentrated on discs with a charged-particle
composition that can be represented as a two-component plasma consisting
of electrons and comparatively massive ions, with the parameter $q$
(equation~\ref{eq:q}) being $\ll 1$. This composition is appropriate to
relatively low-density disc regions or even to denser zones that do not
contain dust (see footonote~\ref{fn:dust}). Weakly ionized disc regions
that have densities $\gtrsim 10^{-13}\, {\rm g\, cm}^{-3}$ and contain
dust may be better approximated by a two-component, $q=1$ plasma
corresponding to oppositely charged grains of equal mass (see
Section~\ref{subsec:model}). In this case the Hall regime is not present
(see equation~\ref{eq:sigmaratio1}) and the parameter space of viable
solutions consists of only two regimes (ambipolar diffusion and
Ohm). One can readily verify that, under these circumstances, the Ohm
domain contains only two sub-regimes, the analogues of the Ohm
Cases~($i$) and~($ii$) for a $q\ll 1$ plasma.\footnote{Note from
equations~(\ref{eq:sigmaratio1})-(\ref{eq:sigmaratio3}) that the
parameter $\psi=\Upsilon_0$ always appears in the conductivity tensor in
the combination $(1+q)\,\psi$. Therefore, the parameter $\Upsilon_0$ in
the expressions for the $q\ll 1$ Cases~($i$) and~($ii$) in
Tables~\ref{table:Ohm_constraints} and~\ref{table:Ohm_constraints1}
is replaced by $2\Upsilon_0$ in the corresponding expressions for
$q=1$.} This finding is not surprising in view of the fact (see
Section~\ref{subsec:Ohm_conductivity}) that the Ohm Case~($iii$) for a
$q\ll 1$ plasma arises from the Hall-current term in Ohm's law, which is
missing when $q=1$.

The viability constraints that we derived on diffusive accretion discs
in which the entire excess angular momentum is transported by a
centrifugally driven wind can be useful also for generalizing the model
to incorporate other modes of angular momentum transport. In particular,
one can argue \citep[see][]{SKW07} that radial angular-momentum
transport through MRI-induced turbulence could operate at disc locations
where the sub-Keplerian rotation constraint (expressed by the first
inequality in Tables~\ref{table:constraints}
and~\ref{table:Ohm_constraints}, in which the various parameters are
taken to have their {\em local}, density-dependent values) is
violated. For example, in the ambipolar diffusion case (as well as in
the Hall Cases~$i$ and~$ii$ and the Ohm Cases~$i$ and~$ii$) this
corresponds to $2 \Upsilon a^2 \propto \rho_{\rm i}/\rho$ being $<1$.  As
this quantity is expected to increase with height on account of the
increase in the fractional ionization of the gas on going away from the
disc midplane, one can envision a situation in which $2\Upsilon a^2$ is $<1$
(with both radial and vertical angular-momentum transport taking place)
in the vicinity of the midplane but becomes $> 1$ (with only vertical
transfer into a wind remaining relevant) closer to the disc
surfaces. This criterion can be employed for constructing a ``hybrid''
disc model in which both vertical and radial angular momentum transport
take place (through a large-scale, ordered magnetic field and a
small-scale, disordered field, respectively; see \citealt{SKW07} for
details).

\section{Conclusion}
\label{sec:conclude}

We have investigated a protostellar accretion disc model in which the
dominant angular momentum transport mechanism is a centrifugally driven
wind launched along a large-scale, ordered magnetic field that threads
the disc. The basic assumptions we adopted are the same as those of the
radially localized disc model constructed by WK93, and our main purpose
has been to extend the analysis presented in that paper, which focussed
on discs in which ambipolar diffusion dominates the magnetic diffusivity
inside the disc, to a general diffusivity regime. In particular, we
included also the Hall and Ohm diffusivities, which are relevant to
higher-density disc regions than those covered by the WK93 treatment. We
employed a tensor-conductivity scheme that, in conjunction with an
ionization-balance calculation, can be used to determine realistic
vertical conductivity profiles for protostellar discs in the context of
this model. However, in this paper and its follow-up, Paper~II, we adopt
a simpler treatment and assume that the three model parameters that are
needed to identify the distinct solution regimes for an ion--electron
plasma remain constant with height in the disc. The three variables
employed for this purpose in the current paper are the electron and ion
Hall parameters ($\beta_{\rm e}$ and $\beta_{\rm i}$, respectively, with
$|\beta_{\rm i}| \ll |\beta_{\rm e}|$) and the neutral--ion coupling
parameter $\Upsilon$ (equation~\ref{eq:eta}). In Paper~II we use instead
two independent ratios of the conductivity tensor components as well as
the Elsasser number $\Lambda$ (see items ($iii$) and ($iv$) in
Section~\ref{subsec:param}).

We determined the parameter regimes where physically viable disc
solutions could be found by employing the hydrostatic approximation and
imposing the following requirements (originally applied in the ambipolar
diffusion regime by WK93 and \citealt{Kon97}) on ``strongly coupled''
systems (discs in which the magnetic field is dynamically well coupled
to the bulk of the matter already at the midplane): (i) the disc (in
contradistinction to the wind) rotates at sub-Keplerian speeds; (ii) a
wind launching condition (which yields a lower limit on $B_r/B_z$) is
satisfied at the disc surface; (iii) most of the disc material is
located below the wind-launching region; and (iv) the rate of Joule
dissipation does not exceed the rate of gravitational energy release in
the disc. Our main results can be summarized as follows:
\begin{itemize}
\item In addition to the ambipolar diffusion regime considered in WK93,
there are four distinct sub-regimes in the Hall diffusion-dominated
parameter domain and three distinct sub-regimes in the Ohm domain. (A
fourth potential Ohm sub-regime was found not to be consistent with the
wind-launching requirement.) The four Hall sub-regimes naturally divide
into two ambipolar diffusion-like and two Ohm-like parameter regions: In
the former pair one of the sub-regimes has the same structural
properties as the ambipolar diffusion regime, whereas in the latter pair
the solution characteristics of one of the sub-regimes are identical to
those of one of the Ohm sub-regimes. In the case (not treated in detail
in this paper) of a high-density disc plasma dominated by oppositely
charged grains of equal mass, only an ambipolar diffusion regime and an
Ohm regime (with two sub-regimes corresponding to our Ohm Cases~$i$
and~$ii$) are present.

\item All viable solutions, irrespective of the diffusivity regime, are
found to satisfy $\Upsilon \gtrsim 1$. The physical requirement
expressed by this condition, that the neutral--ion momentum exchange
time be shorter than the disc orbital time, is thus indicated to be a
fundamental constraint on wind-driving discs of this type.

\item Viable solutions are also characterized by the parameters $a_0$
(midplane Alfv\'en-to-sound speed ratio) and $\epsilon$ (midplane
inflow-to-sound speed ratio) being $\lesssim 1$. However, weakly coupled
discs, characterized by values of $\Lambda_0$ (the midplane value of the
Elsasser number; see Section~\ref{subsec:param}) that are $\ll 1$, could
have $a_0\ll 1$ and $\epsilon\ll 1$. In such systems a significant
bending of the magnetic field away from the rotation axis (required for
driving a centrifugally driven wind) takes place only above some finite
height in the disc, although vertical angular momentum transport along
the large-scale field typically occurs already at lower elevations.
Although our analysis does not directly apply to weakly coupled discs,
our results should be useful for understanding the basic qualitative
aspects of their behaviour.

\item The wind-launching and wind-loading conditions (the aforementioned
requirements~($ii$) and~($iii$), respectively) together imply that
magnetic squeezing (by the gradient of the magnetic pressure force
associated with the transverse field components $B_r$ and $B_\phi$)
dominates the gravitational tidal compression of the disc. A more
elaborate model is, however, needed to verify the full generality of
this result.

\item The transverse magnetic field at the disc surface is dominated by
the radial component in all the Hall parameter sub-regimes (including
their ``twins'' in the ambipolar-diffusion and Ohm domains), and by the
azimuthal component in the remaining two Ohm sub-regimes. However, even
in the latter two cases the poloidal surface magnetic field must be
sufficiently strongly bent to satisfy the wind launching requirement.
\end{itemize} 

Centrifugally driven winds are not the only means of vertical angular
momentum transport in a protostellar disc (magnetic braking is an
example of another option), and there are also alternative mechanisms
that involve radial transport along the plane of the disc. It is
nevertheless instructive to investigate the limiting case in which wind
transport dominates at a given radius in view of the ubiquity
of energetic outflows in protostellar systems and the inferred
association of disc winds with FU Orionis-type high-accretion-rate
events. The constraints considered in our model can also be useful for
identifying the parameter regimes where other mechanisms (such as
MRI-induced turbulence) are likely to operate, possibly even at the same
radial location as wind-mediated angular momentum transport
\citep{SKW07}.

The parameter regimes that we identified were determined under the
assumption that the values of the parameters used in the classification
scheme did not vary with height in the disc. This approximation becomes
less accurate as one moves from the ambipolar-diffusion regime to the
Hall regime and then (at even higher disc densities and column
densities) to the Ohm regime. In particular, in the latter case it can
be expected that if a real disc could be described as being in this
regime at the midplane then it would transit to the Hall regime further
up and would likely be ambipolar diffusion-dominated at the surface. For
this reason (and also because wind-driving discs might not be massive
enough to harbour an Ohm regime within their weakly ionized zones) we
have emphasized the Hall regime in our discussion and do not consider
the Ohm domain in Paper~II. However, our findings for the Ohm regime
could still be applicable to the innermost, collisionally ionized
regions of protostellar discs (and possibly also to disc regions
participating in an FU Orionis-type outburst), in which an anomalous
ohmic resistivity might dominate.

The actual parameter regime that characterizes a given section of an
accretion disc of the type that we have modelled will be determined by
the global magnetic flux distribution, the density structure (which
depends in part on the mass accretion rate) and the ionization profile
(which depends in part on the nature of the ionization sources as well
as on the disc composition). It is conceivable that some of the
sub-regimes that we have identified are not commonly realized in Nature
or that they correspond to unstable configurations. More insight into
these questions would likely emerge from comparisons with observations
as well as from additional analytic and numerical work. Our study has,
however, indicated that, if such wind-driving accretion flows are indeed
present in real protostellar systems, they will exhibit certain generic
properties (including good neutral--ion coupling, subthermal Alfv\'en
and inflow speeds and magnetic squeezing within the field-line bending
region) irrespective of the particular parameter regime to which they
correspond.

\section*{Acknowledgments}

We thank the reviewer for helping us improve the presentation of this
paper. This research was supported in part by NASA Theoretical
Astrophysics Programme grant NNG04G178G (AK and RS), by NSF grant
AST-0908184 (AK) and by the Australian Research Council grants DP0344961
and DP0881066 (RS and MW).

\appendix
\section{The $\epsilon_{\rm B} \approx 0$ approximation}
\label{sec:appA}
In their consideration of a weakly ionized disc containing a
two-component plasma, WK93 noted the qualitative similarity of solutions
characterized by the same value of the parameter combination $(\epsilon -
\epsilon_{\rm B})$. They attributed this result to the fact that the only
change in the underlying system of equations introduced by switching to
a reference frame that moves with the radial velocity $v_{{\rm B} r 0} = -
\epsilon_{\rm B} c_{\rm s}$ of the poloidal flux surfaces involves the
magnitude of the torque that is required to remove the excess angular
momentum, so that all radial-velocity terms (except in the angular
momentum equation, which remains unchanged) are modified simply by subtracting
$v_{{\rm B} r 0}$ from the radial velocity component. We appealed to
this result in Section~\ref{subsec:param} to motivate setting
$\epsilon_{\rm B} = 0$ in our analysis.

One could however question this approach on the basis of the following
consideration. Starting from Ohm's law in the form $\E' = \E + {\v
\cross \B}/{c} = (4\pi/c^2) \bmath{\eta}\bdot \J$, where $\bmath{\eta}$
is the diffusivity tensor, we rewrite it in terms of normalized
quantities as 
\begin{equation} 
\left( \begin{array}{c}
w_{{\rm B} r 0} - w_r \\ 
\\ 
\w_{{\rm B} \phi 0} - w_\phi 
\end{array} \right ) = \left ( \begin{array}{cc} 
\tilde{\eta}_{\rm H} & - \tilde{\eta}_{\rm P} \\ 
\\ 
\tilde{\eta}_{\rm P} & \tilde{\eta}_{\rm H} 
\end{array} \right) \left( \begin{array}{c} 
0.5\,w_r\\ 
\\ 
2\,w_\phi 
\end{array} \right) \ ,
\label{App_eq1:matrix} 
\end{equation} 
\vspace{10pt} 

\noindent
where $\tilde{\eta}_{\rm P} = \tilde{\eta}_{\rm O} + \tilde{\eta}_{\rm
A}$ and where $\tilde{\eta}_{\rm O}$, $\tilde{\eta}_{\rm H}$ and
$\tilde{\eta}_{\rm A}$ are obtained by multiplying the expressions in
equations~(\ref{eq:etaO})--(\ref{eq:etaA}), respectively, by $1/h_{\rm
T}c_{\rm s}a_0^2$. We have also written $w_{{\rm B} r 0} \equiv v_{{\rm
B} r 0}/c_{\rm s} = w_{{\rm E} \phi 0}$ and $w_{{\rm B} \phi 0} \equiv
r(\Omega_{{\rm B} 0} - \Omega_{\rm K})/c_{\rm s} = - w_{{\rm E} r 0}$
for the normalized midplane radial and azimuthal velocities,
respectively, of the poloidal flux surfaces. By inspecting
equation~(\ref{App_eq1:matrix}) one may be induced to infer that, since
the off-diagonal terms of the diffusivity tensor dominate in the
ambipolar-diffusion and Ohm regimes whereas the diagonal terms are
dominant in the Hall limit, one should retain the latter terms when
considering the Hall regime and the former terms when treating the other
limits. This inference could be carried over to the analysis of
equation~(\ref{eq1:matrix}) in the text, in which the matrix is related
to the inverse of the matrix that appears in
equation~(\ref{App_eq1:matrix}). Specifically, specializing to the
midplane (where $b_r = b_\phi = 0$) and assuming $q \ll 1$ for
simplicity, one can use
equations~(\ref{eq:sigmaratio1})--(\ref{eq:sigmaratio3}) to deduce that
$A_1 = - A_4 \propto \sigma_{\rm P}/\sigma_\perp$ (with the factor
$q\beta^2$ in equations~\ref{eq:A1} and~\ref{eq:A4} representing the
ratio $\eta_{\rm O}/\eta_{\rm A}$). One can similarly deduce that the
term $\beta$ in the expressions for $A_2$ and $A_3$ is $\propto
\sigma_{\rm H}/\sigma_\perp$ (with the term involving $\psi$ in
equations~\ref{eq:A2} and~\ref{eq:A3} arising from the advective
contribution to the electric field). Adopting the above reasoning, it
would appear that one should retain the diagonal terms $A_1$ and $A_4$
in the ambipolar-diffusion and Ohm regimes and the off-diagonal terms
$A_2$ and $A_3$ in the Hall limit. However, this procedure cannot be
implemented if one sets $\epsilon_{\rm B}$ to be identically zero in
equation~(\ref{eq1:matrix}), as we have done in our analysis, which
casts doubt on the validity of the approach followed in the text.

In addressing this question it is important to keep in mind the basic
attributes of the system that we wish to model. We are primarily
interested in the quasi-steady behaviour of a disc in which mass and
poloidal magnetic flux that were originally part of a large-scale
equilibrium configuration of a molecular cloud core are carried in by
the accretion flow. In this picture, the accretion is enabled in large
measure by the vertical magnetic transport of angular momentum, and the
inward motion of the matter is, in turn, responsible for the radial
transport of the field lines (which, however, generally lag behind the
matter because of the disc's diffusivity). The vertical magnetic transport
of angular momentum implies that $J_rB_z>0$ and hence that $w_r < 0$
(see equation~\ref{eq:phi_alg}), whereas the inward bending of the
magnetic field lines by the inflowing gas implies that $J_\phi B_z> 0$
and hence that $w_\phi < 0$ (see equation~\ref{eq:r_alg}). The fact that
the motion of the field lines is controlled by that of the gas implies
that $w_{{\rm B} \phi}= -w_{{\rm E}r}$ would also be $< 0$ (see
Section~\ref{subsec:dimensionless}). In a similar vein, we expect the
radial velocity of the poloidal field lines (described by the parameter
$w_{{\rm B} r 0} = - \epsilon_{\rm B}$) to satisfy $w_{r 0} \le w_{{\rm
B} r 0}$ (or, equivalently, $\epsilon_{\rm B} \le
\epsilon$).\footnote{It is conceivable that $\epsilon_{\rm B}$ and
$\epsilon$ could have opposite signs, but such a situation would likely
only arise in the context of a localized, time-dependent phenomenon. An
example of this possibility is provided by the cyclical behaviour
exhibited by the core-collapse solutions of \citet{TM05} near the
boundary of their computational ``central sink.''  This behaviour is
indeed localized (in both space and time) and clearly does not
correspond to the global, quasi-steady accretion process analyzed in
this paper.}

We now proceed to examine the consequences of the alternative
formulation outlined above in light of the just-discussed
constraints. To this end, we substitute for $j_{r 0}$ and $j_{\phi 0}$
in equation~(\ref{eq1:matrix}) from equations~(\ref{eq:phi_alg})
and~(\ref{eq:r_alg}), respectively, to obtain
\begin{equation}
A_1\, w_{{\rm E} r 0} + A_2\, \epsilon_{\rm _B} = \frac{K \epsilon}{2\psi}\ ,
\label{App_eq:A1A2}
\end{equation}
\begin{equation}
A_3\, w_{{\rm E} r 0} + A_4 \, \epsilon_{\rm B} = -\frac{2 K w_{\phi
    0}}{\psi}\ .
\label{App_eq:A3A4}
\end{equation}
In the ambipolar-diffusion and Ohm regimes we retain only the $A_1$ and
$A_4$ terms. In the ambipolar diffusion limit $A_1 = -A_4 \approx 1$ and
$K \approx 1 + \psi^2 \approx \psi^2$ (since $\psi$ is equal in this
case to the midplane value of the neutral--field coupling parameter
$\Lambda$, which is expected to be $> 1$), so
equation~(\ref{App_eq:A3A4}) implies $\epsilon_{\rm B} = 2 \psi w_{\phi
0} < 0$, which is inconsistent with our expectation that $\epsilon _B
\ge 0$. Similarly, in the Ohm limit $A_1 = -A_4 \approx q\beta^2$ and $K
\approx q\beta^2 + \psi^2$, so equation~(\ref{App_eq:A3A4}) implies
$\epsilon_{\rm B} = 2 (\psi/q\beta^2 + q\beta^2/\psi) w_{\phi 0}$, which
is again inferred to be $< 0$, contrary to our physical expectation. In
a similar vein, we retain only the $A_2$ and $A_3$ terms in the Hall
regime. Using equations~(\ref{eq:A2}) and~(\ref{eq:A3}) and focussing on
the case $q\ll 1$, we infer from equations~(\ref{App_eq:A1A2})
and~(\ref{App_eq:A3A4}) that
\begin{equation}
\frac{\epsilon_{\rm B}}{\epsilon} = \frac{1 +
\frac{5}{2}\frac{\psi}{\beta}+\left( \frac{\beta}{\psi}\right )^2}{1 +
\frac{2\beta}{\psi}}
\label{App_eq:epsB_eps}
\end{equation}
and
\begin{equation}
\frac{w_{{\rm E} r 0}}{-w_{\phi 0}} = \frac{1 +
\frac{5}{2}\frac{\psi}{\beta}+\left( \frac{\beta}{\psi}\right )^2}{1 +
\frac{\beta}{2\psi}}\ .
\label{App_eq:wEr_wphi}
\end{equation}
One can readily verify that equations~(\ref{App_eq:epsB_eps})
and~(\ref{App_eq:wEr_wphi}) cannot be simultaneously satisfied while
also respecting the constraints $0\le \epsilon_{\rm B}/\epsilon \le 1$ and 
$w_{{\rm E} r 0}/w_{\phi 0}<0$. These arguments indicate that a formulation
based simply on the dominant terms in the conductivity tensor is not
physically self-consistent for the problem under
consideration.

In contrast, one can verify that the formulation adopted in the text is
self-consistent -- in the sense that the neglected terms in
equations~(\ref{App_eq:A1A2}) and~(\ref{App_eq:A3A4}) never come to
dominate the ones that were retained -- under the assumption
$\epsilon_{\rm B}/\epsilon \le 1$ . In particular, using the
expression~(\ref{eq:er}) for $w_{{\rm E}r0}$ and
equations~(\ref{eq:A1})--(\ref{eq:A4}), it is straightforward to ascertain
that the ratios $|A_2\epsilon_{\rm B}/A_1 w_{{\rm E}r0}|$ and
$|A_4\epsilon_{\rm B}/A_3 w_{{\rm E}r0}|$ remain $\le 1$ in this case
for all the viable-solution parameter regimes identified in
Sections~\ref{sec:constraints} and~\ref{sec:Ohm_constraints}.

\bsp 
\label{lastpage}
\end{document}